\documentclass[prd,aps,reprint,onecolumn,nofootinbib]{revtex4-1}
\usepackage{graphicx,amsmath,amssymb,graphics}
\usepackage{hyperref}
\hypersetup{colorlinks = true, citecolor = blue}
\usepackage{vector}

\begin{document}

\title{Gauge invariant cosmological perturbations for the nonminimally coupled inflaton field}

\author{Jan Weenink}
\email{j.g.weenink@uu.nl}
\author{Tomislav Prokopec}
\email{t.prokopec@uu.nl}
\affiliation{Institute for Theoretical Physics and Spinoza Institute,\\
Utrecht University, Leuvenlaan 4, 3585 CE Utrecht, The Netherlands}

\begin{abstract}
We construct the gauge invariant free action for cosmological perturbations for the nonminimally coupled inflaton field in the Jordan frame. For this the phase space formalism is used, which keeps track of all the dynamical and constraint fields. We perform explicit conformal transformations to demonstrate the physical equivalence between the Jordan and Einstein frames at the level of quadratic perturbations. We show how to generalize the formalism to the case of a more complicated scalar sector with an internal symmetry, such as Higgs inflation. This work represents a first step in developing gauge invariant perturbation theory for nonminimally coupled inflationary models.
\end{abstract}

\maketitle

\section{Introduction}
Models of inflation \cite{Guth:1980} considering a nonzero coupling $\xi$ between the scalar inflaton field and the gravitational Ricci scalar have been studied for more than 20 years.
Originally introduced to solve the graceful exit problem in the old inflationary scenario \cite{La:1989za}, it was soon realized that a nonminimal coupling could also improve the chaotic inflationary scenario \cite{Salopek:1988qh,Futamase:1987ua,Fakir:1990eg}. We will motivate the use of a nonzero $\xi$ later on. In the context of inflation a nonminimal coupling $\xi$ can not only relax the initial conditions for chaotic inflation, but it can also weaken the constraints on the inflaton potential. For successful chaotic inflation the quartic self-coupling of a minimally coupled inflaton field should take the unnaturally small value $\lambda \sim 10^{-13}$, but a (large) nonminimal coupling $\xi$ modifies this condition as $\lambda/|\xi|^2\sim 10^{-13}$. Therefore $\lambda$ can increase by many orders of magnitude if $|\xi|$ is sufficiently large. This allows the Higgs boson itself to be the inflaton field \cite{Salopek:1988qh}, an appealing idea that was revived recently by Bezrukov and Shaposhnikov \cite{Bezrukov:2007ep}. \\
The constraints on the inflaton potential are obtained from observations of the temperature fluctuations in the CMB\cite{Komatsu:2010fb}. The temperature fluctuations are ultimately a consequence of quantum fluctuations of the inflaton field, which are in turn described by the theory of gauge invariant cosmological perturbations \cite{Bardeen:1980kt,Mukhanov:1981xt,Hawking:1982cz,Starobinsky:1982ee,Guth:1982ec,Bardeen:1983qw}. Scalar perturbations of the metric and inflaton field, coupled through the Einstein equations, beautifully combine into a single gauge invariant variable often referred to as the comoving curvature perturbation.\\
At first the theory of cosmological perturbations was derived for a minimally coupled scalar field. However, Mukhanov et al. \cite{Mukhanov:1990me} correctly pointed out that a nonminimal coupling of the inflaton field to gravity can be removed by performing a conformal transformation of the metric $g_{\mu\nu,E}=\omega^2 g_{\mu\nu}$. Therefore it is in principle sufficient to know the standard results (\textit{e.g.} the primordial power spectrum) in the frame where the scalar-gravity coupling is minimal, the so-called Einstein frame. The Jordan frame (\text{i.e.} nonminimal coupling) results are then obtained by performing the conformal transformation. Although the Jordan and Einstein frame are physically equivalent at the classical level, it is not obvious that the frames are also equivalent at the level of (quantum) fluctuations. However, Makino and Sasaki \cite{Makino:1991sg} and Fakir et al. \cite{Fakir:1992cg} proved that the comoving curvature perturbation is not only gauge invariant, but also invariant under a conformal transformation. This means that it is for example possible to calculate the action for the comoving curvature perturbation in the Einstein frame and then obtain the action in the Jordan frame by performing a conformal transformation of the metric, which was done by Hwang \cite{Hwang:1996np}. Before, Hwang and Noh\cite{Hwang:1996xh} already found the field equation for the comoving curvature perturbation in the Jordan frame, by considering the linearized Einstein equations for a nonminimally coupled scalar field in the uniform field and curvature gauges.
\\
In this paper it is our goal to derive the gauge invariant free action for the nonminimally coupled inflaton field. To avoid any confusion regarding gauge freedom or conformal invariance between Jordan and Einstein frames, we do not fix a gauge or perform a conformal transformation in the derivation. Instead we will keep using all dynamical and constraint fields in the action and work exclusively in the Jordan frame. As we will see we obtain in a straightforward and unambiguous way the completely gauge invariant action for the nonminimally coupled scalar field, where as expected the only dynamical degrees of freedom are the (scalar) comoving curvature perturbation and the (tensor) graviton. We will use the canonical approach put forward recently in Ref. \cite{Prokopec:2010be}. This new approach is fundamental and very general since it keeps all the constraint fields and can in principle be generalized to arbitrary order in field perturbations. Moreover we will perform our calculations in $D$ dimensions, anticipating dimensional regularization in future loop calculations.\\
Our work is motivated by a number of points. First of all the gauge invariant action for cosmological perturbations is crucial in order to calculate quantum corrections to the inflaton potential. The quantum corrected inflaton potential determines whether or not the conditions for slow-roll inflation are met. In this paper we consider a nonminimally coupled inflaton field and we show that we can consistently calculate the free action. The next step is to derive the higher order gauge invariant action, which is outlined in \cite{Prokopec:2010be}. A second motivation for our work is to establish the physical equivalence of the Jordan and Einstein frames at the level of the free action. The main complication in this respect is the fact that the Einstein and Jordan frames are related by nonlinear field transformations. In fact we will show that the two frames are also physically equivalent when considering field fluctuations up to quadratic order. Chisholm \cite{Chisholm1961469} and Kamefuchi et al. \cite{Kamefuchi1961529} already proved almost 50 years ago that, although the field equations may differ in detail under point-transformations of the fields (\text{i.e.} transformations without time derivatives of fields), the (Euler-Lagrange) form of these equations and of the stress-energy tensor remains identical. Thereby the (quantum) equivalence of two frames related by nonlinear field transformations is established. In this paper we would like to understand how this equivalence works in detail for the case of the conformal transformation when applied to Einstein's gravity coupled nonminimally to a scalar matter.\\
As for the motivation of the use of a nonzero nonminimal coupling $\xi$: if $\xi$ does not have the conformal coupling $\frac16$ (which is the case), then $\xi$ will run with the energy scale, see Ref. \cite{Bilandzic:2007nb}. In other words, if we pick $\xi$ to be zero at some scale, it will not be zero at another energy scale. Moreover, if we pick $\xi$ to be large at some scale, then $\xi$ will remain large since the running is generically logarithmic. In the end of course it is Nature who decides which value $\xi$ takes at some energy scale. Fortunately, if $\xi$ would be nonzero then we should in principle be able to observe this. Minimal and nonminimal inflationary models are physically different since the matter and gravitational fields propagate differently if $\xi$ is nonzero. This is true in both the Einstein and Jordan frame. \\
The outline of the paper is the following: in section \ref{sec:canonicalaction} we formulate the action for the nonminimally coupled inflaton field in canonical form. We derive the background Friedmann and field equations and show the relation to the background fields in the Einstein frame. In section \ref{sec:freeaction} we perturb the action up to quadratic order in field fluctuations and perform a diagonalization procedure of this action. Our final and most important result is the completely gauge invariant free action for the nonminimally coupled inflaton field. By performing the conformal transformation we show that this action, including both dynamical and constraint fields, is physically equivalent to the quadratic action in the Einstein frame. Finally in section \ref{sec:Higgsinflation} we generalize our result to the case of Higgs inflation, where the Higgs boson itself is the inflaton field. We briefly discuss the idea and the current status of Higgs inflation and show that a scalar field theory with a local $SU(N)$ or $O(N)$ symmetry contains one dynamical inflaton field.

{ \section{\label{sec:canonicalaction}Canonical action for the nonminimally coupled inflaton field} }
We start with the $D$-dimensional action for a scalar field $\Phi$ that is coupled to the Ricci scalar $R$ through some function $F(\Phi)$,
\begin{equation}
S=\int d^D x \sqrt{-g}\left\{-R^{(D)} F(\Phi)-\frac12 g^{\mu\nu}\partial_{\mu}\Phi\partial_{\nu}\Phi -V(\Phi)\right\}.
\label{nonminimalaction}
\end{equation}
The metric convention is $(-,+,+,+)$ and we will work in units where $\hbar=c=1$. For a nonminimally coupled inflaton field $F(\Phi)=\frac12 M_P^2 -\frac12 \xi \Phi^2$ (where $M_P^2=(8\pi G_N)^{-1}$), where $\xi=+\frac{D-2}{4(D-1)}$ is the conformal coupling value and $\xi=0$ corresponds to minimal coupling. In the following we will keep $F(\Phi)$ completely general. To avoid any confusion for the rest of this paper we label the $D$-dimensional Ricci scalar with an index $D$.\\
In the Lagrange formulation the action \eqref{nonminimalaction} is invariant under coordinate transformations of the metric field $g_{\mu\nu}$. It is precisely this coordinate invariance however which makes the extraction of true dynamical fields problematic. Because we are interested in the dynamical fields in the context of cosmological perturbations, we therefore want to break the general covariance of the metric by separating spacetime into spatial surfaces of constant time. To this end we use the ADM\cite{Arnowitt:1962hi} decomposition of the metric with the line element
\begin{equation}
ds^2=-N^2dt^2+g_{ij}(dx^i+N^idt)(dx^j+N^jdt),
\label{ADMlineelement}
\end{equation}
where $N$ and $N^i$ are called the lapse and shift functions respectively. Under a time change $dt$ the corresponding change in a coordinate $x^i$ is $Ndt$ in the direction perpendicular to the spatial surface, and $N^idt$ in the direction parallel to the surface. This geometrical interpretation shows that the lapse and shift functions correspond to coordinate changes, which seems to leave the spatial metric $g_{ij}$ as the true dynamical field. In fact we can determine this precisely in the Hamiltonian formulation of gravity, which is obtained using the ADM metric. The ADM formalism is therefore necessary for a first principle quantization and can be used to check the correctness of any other quantization procedure. After a series of steps (presented in Appendix \ref{secappendixADMaction}) where we derive the canonical momenta and substitute these back into the action, we obtain the action for a nonminimally coupled scalar field in canonical form,
\begin{equation}
S=\int d^D x \left[p^{ij}\partial_0 g_{ij}+p_{\Phi}\partial_0{\Phi}-N\mathcal{H}-N_i\mathcal{H}^{i}\right],
\label{canonicalactionnonminimal}
\end{equation}
where
\begin{align}
\nonumber \mathcal{H}=&-\sqrt{g}RF+\frac{1}{\sqrt{g} F}\left[p^{ij}g_{ik}g_{jl}p^{kl}-\frac{1}{D-2}\frac{\left(1+2\frac{F^{\prime 2}}{F}\right)}{\Omega}p^2\right]\\
&+\frac{1}{\sqrt{g}}\frac{1}{2\Omega}p_{\Phi}^2+\sqrt{g}\frac{1}{2}g^{ij}\partial_i\Phi\partial_j\Phi+\sqrt{g}V(\Phi)
-\frac{1}{\sqrt{g}F}\frac{2}{D-2}\frac{1}{\Omega}F'p p_{\Phi}+2\sqrt{g}g^{ij}\nabla_i\nabla_jF\label{Nconstraint}\\
\mathcal{H}^{i}=&\partial^i\Phi p_{\Phi}-2\nabla_jp^{ij}\label{Niconstraint}.
\end{align}
$p^{ij}$ and $p_{\Phi}$ are the canonical momenta conjugate to $g_{ij}$ and $\Phi$ respectively and $p\equiv g_{ij}p^{ij}$. The action \eqref{canonicalactionnonminimal} is a new result and indeed reduces to the well known canonical action for a minimally coupled scalar field if we set $F=\frac12 M_p^2\equiv 1$. The canonical action indeed shows that the only dynamical field is $g_{ij}$, whereas the lapse $N$ and shift $N^i$ functions appear as Lagrange multipliers to the constraints. Since $p^{ij}$ is a densitized tensor the covariant derivative is understood as $\nabla_j p^{\ij}=\partial_j p^{ij}+\Gamma^{i}_{jl} p^{jl}$, where $\Gamma^{i}_{jl}$ only depends on spatial derivatives of the spatial metric $g_{ij}$. The Ricci scalar $R$ in \eqref{Nconstraint} is the 'spatial' Ricci scalar and only depends on (spatial derivatives of) $\Gamma^{i}_{jl}$. In the canonical action indices are raised and lowered by the spatial metric $g_{ij}$. Furthermore we use shorthand notation where $F=F(\Phi)$ and $F'=dF/d\Phi$, and we define the convenient variable
\begin{equation}
\Omega= 1+2\frac{D-1}{D-2}\frac{F^{\prime 2 }}{F}.\label{definitionOmega}
\end{equation}
As a consequence of the nonminimal coupling between $\Phi$ and $R$, the latter containing double derivatives, the momenta $p$ and $p_{\Phi}$ are coupled in the Hamiltonian $\mathcal{H}$ in Eq. \eqref{Nconstraint}. Since this leads to coupled equations when we derive the Hamilton equations for $p^{ij}$ and $p_{\Phi}$ we would like to decouple the momenta. We do this by introducing a shifted momentum
\begin{align}
\hat{p}_{\Phi}&\equiv p_{\Phi}-\frac{2}{D-2}\frac{F'}{F}p.
\label{nonminimalscalarmomentumshifted2}
\end{align}
Since the shift in the momentum only depends on $F(\Phi)$ and $p$, the transformation is canonical, thus the resulting Hamilton equations of motion will be equivalent for either $\hat{p}_{\Phi}$ and $p_{\Phi}$. In terms of the shifted momentum $\hat{p}_{\Phi}$ we find that we can write the action \eqref{nonminimalaction} as
\begin{equation}
S=\int d^D x \left[p^{ij}\partial_0 g_{ij}+\hat{p}_{\Phi}\partial_0{\Phi}+\frac{2}{D-2}\frac{F'}{F}p\partial_0\Phi -N\mathcal{H}-N_i\mathcal{H}^{i}\right],
\label{canonicalactionnonminimalshifted2}
\end{equation}
where
\begin{align}
\nonumber \mathcal{H}=&-\sqrt{g}RF+\frac{1}{\sqrt{g} F}\left[p^{ij}g_{ik}g_{jl}p^{kl}-\frac{1}{D-2}p^2\right]\\
&+\frac{1}{\sqrt{g}}\frac{1}{\Omega}\frac{1}{2}\hat{p}_{\Phi}^2+\sqrt{g}\frac{1}{2}g^{ij}\partial_i\Phi\partial_j\Phi +\sqrt{g}V(\Phi)
+2\sqrt{g}g^{ij}\nabla_i\nabla_jF\label{Nconstraintshifted2}\\
\mathcal{H}^{i}=&\partial^i\Phi (\hat{p}_{\Phi}+\frac{2}{D-2}\frac{F'}{F}p)-2\nabla_jp^{ij}\label{Niconstraintshifted2}.
\end{align}
The Hamiltonian $\mathcal{H}$ has dramatically simplified because of the shifted momentum. On the other hand, there are additional terms in the kinetic part of the action \eqref{canonicalactionnonminimalshifted2} and the momentum density $\mathcal{H}^{i}$. Our goal is to perturb the action up to second order in fluctuations around a FLRW background. Therefore we  separate all fields in a classical background plus a small perturbation as
\begin{align}
p^{ij}&=\frac{\mathcal{P}(t)}{2(D-1) a(t)}\left(\delta^{ij}+\pi^{ij}(t,\bvec{x})\right)\label{perturbedpij}\\
\hat{p}_{\Phi}&=\hat{\mathcal{P}}_{\phi}(t)\left(1+\hat{\pi}_{\varphi}(t,\bvec{x})\right)\\
g_{ij}&=a(t)^2\left(\delta_{ij}+h_{ij}(t,\bvec{x})\right)\label{perturbedgij}\\
\Phi&=\phi(t)+\varphi(t,\bvec{x})\\
N&=\bar{N}(t)+n(t,\bvec{x})\label{perturbedN}.
\end{align}
The shift $N^i$ is a pure fluctuation, \textit{i.e.} its background value is zero. Note that we keep working with hatted quantities $\hat{\mathcal{P}}_{\phi}$ and $\hat{\pi}_{\varphi}$ to clarify that these are not the canonical momenta conjugate to $\phi$ and $\varphi$.

\subsection{Background equations}
To recover the background equations we can perturb the action \eqref{canonicalactionnonminimalshifted2} up to linear order in perturbations \eqref{perturbedpij}-\eqref{perturbedN} and set the resulting expressions to vanish. In general this gives the Hamilton equations of motion, which are derived from the background action
\begin{equation}
S^{(0)}=\int d^Dx \left\{\mathcal{P}\partial_0 a+\hat{\mathcal{P}}_{\phi}\partial_0\phi+\frac{1}{D-2}\frac{F'}{F}a\mathcal{P}\partial_0 \phi-\bar{N}\mathcal{H}^{(0)}\right\},
\label{backgroundactionshift2}
\end{equation}
where
\begin{equation}
\mathcal{H}^{(0)}=-\frac{1}{F}\frac{1}{a^{D-3}}\frac{\mathcal{P}^2}{4(D-1)(D-2)}+\frac{\hat{\mathcal{P}}_{\phi}^2}{2\bar{\Omega} a^{D-1}}+a^{D-1}V,
\label{backgroundhamiltonianshift2}
\end{equation}
where $F(\phi)$ and $\bar{\Omega}(\phi)$ are functions of the background fields only. Varying this action with respect to $\mathcal{P}$ and $\hat{\mathcal{P}}_{\phi}$ gives
\begin{align}
\mathcal{P}&=-2(D-1)(D-2)F a^{D-2}\left[H+\frac{1}{D-2}\frac{F'}{F}\dot{\phi}\right]\label{Ponshell}\\
\hat{\mathcal{P}}_{\phi}&=\bar{\Omega} a^{D-1}\dot{\phi}\label{Pphionshell},
\end{align}
where a dotted derivative corresponds to $\dot{a}\equiv \bar{N}^{-1} da/dt$ and we have identified the Hubble parameter as $H\equiv \dot{a}/a$. Since $\bar{N}$ can be picked arbitrary, the action \eqref{backgroundactionshift2} is time reparametrization invariant, a remnant of the diffeomorphism invariance of the original action \eqref{nonminimalaction}. Equations \eqref{Ponshell} and \eqref{Pphionshell} are the on-shell expressions for $\mathcal{P}$ and $\hat{\mathcal{P}}_{\phi}$. A variation of the background action \eqref{backgroundactionshift2} with respect to $a$ and $\phi$ gives
\begin{align}
\dot{\mathcal{P}}&=\frac{1}{D-2}\frac{F'}{F}\mathcal{P}\dot{\phi}-\frac{1}{F}\frac{D-3}{a^{D-2}}\frac{\mathcal{P}^2}{4(D-1)(D-2)}+ \frac{D-1}{\bar{\Omega}}\frac{\hat{\mathcal{P}}_{\phi}^2}{2a^{D}}-(D-1)a^{D-2}V
\label{fieldequationP2}\\
\dot{\hat{\mathcal{P}}}_{\phi}&=-\frac{1}{D-2}\frac{F'}{F}\left(a\mathcal{P}\right)^{\cdot}+\left(\frac{1}{F}\right)'\frac{1}{a^{D-3}}\frac{\mathcal{P}^2}{4(D-1)(D-2)}-\left( \frac{1}{\bar{\Omega}}\right)'\frac{\hat{\mathcal{P}}_{\phi}^2}{2a^{D-1}}-a^{D-1}V_{,\phi},
\label{fieldequationphi2}
\end{align}
where $V_{,\phi}=dV/d\phi$. Finally we can vary the background action with respect to $\bar{N}$ to find the constraint equation
\begin{equation}
\frac{1}{F}\frac{1}{a^{D-3}}\frac{\mathcal{P}^2}{4(D-1)(D-2)}=\frac{\hat{\mathcal{P}}_{\phi}^2}{2\bar{\Omega} a^{D-1}}+a^{D-1}V.
\label{constraintonshell}
\end{equation}
If we insert the canonical momenta \eqref{Ponshell} and \eqref{Pphionshell} in Eqs. \eqref{fieldequationP2}-\eqref{constraintonshell} we obtain the background Friedmann and field equations
\begin{align}
&H^2=\frac{1}{(D-1)(D-2)F}\left[\frac12 \dot{\phi}^2+V-2(D-1)H\dot{F}\right]
\label{nonminimalH2}\\
&\dot{H}=\frac{1}{(D-2)F}\left(-\frac12\dot{\phi}^2+H\dot{F}-\ddot{F}\right)
\label{nonminimaldotH}\\
&\ddot{\phi}+(D-1)H\dot{\phi}-(D-1)\left(D H^2+2\dot{H}\right)F'+V_{,\phi}=0,
\label{fieldequation}
\end{align}
where we recognize the background Ricci scalar
\begin{equation}
R=(D-1)\left(D H^2+2\dot{H}\right).
\label{ricciscalarFLRW}
\end{equation}
Eqs. \eqref{nonminimalH2}-\eqref{fieldequation} agree with the Friedmann and field equations obtained from a variation of the action \eqref{nonminimalaction} with respect to $g^{\mu\nu}$ and $\phi$, see for example Ref. \cite{Hwang:1996xh}. Note that in all the above equations the minimal result is recovered by setting $F=\frac12 M_P^2\equiv 1$. Furthermore, only two out of the three equations \eqref{nonminimalH2}-\eqref{fieldequation} are independent.
\subsection{\label{section:classicalequivalence} Classical equivalence of Jordan and Einstein frames}
The Einstein frame is the frame in which the inflaton field is minimally coupled to gravity. In the Einstein frame the action in canonical form is \cite{Prokopec:2010be}
\begin{equation}
S^{(0)}=\int d^Dx \left\{\mathcal{P}_E\partial_0 a_E+\mathcal{P}_{\phi,E}\partial_0\phi_E-\bar{N}_E\mathcal{H}_E^{(0)}\right\},
\label{backgroundactionEinstein}
\end{equation}
where
\begin{equation}
\mathcal{H}_E^{(0)}=-\frac{1}{a_E^{D-3}}\frac{\mathcal{P}_E^2}{4(D-1)(D-2)}+\frac{\mathcal{P}_{\phi,E}^2}{2 a_E^{D-1}}+a_E^{D-1}V_E.
\label{backgroundhamiltonianEinstein}
\end{equation}
The subscript $E$ denotes the quantities in the Einstein frame. This action can be obtained from Eq. \eqref{backgroundactionshift2} by setting $F=\frac12 M_P^2\equiv 1$. We can always make a transformation from the Einstein frame to the Jordan frame (with nonminimal coupling) by performing a conformal transformation of the metric,
\begin{equation}
g_{\mu\nu,E}=\omega^2 g_{\mu\nu},
\label{conformaltransformation}
\end{equation}
where $\omega=\omega\left(\Phi(x)\right)$. It is a well known fact that the Einstein and Jordan frames are physically equivalent at the level of the background equations of motion. Let us now establish this physical equivalence for the background Einstein and Jordan frame actions. Thus, we want to find out how the background fields in the Einstein frame action \eqref{backgroundactionEinstein} should be rescaled in order to arrive at the Jordan frame action \eqref{backgroundactionshift2}. Considering the ADM metric \eqref{ADMlineelement} the background lapse $\bar{N}_E$ and scale factor $a_E$ transform under the conformal transformation \eqref{conformaltransformation} as
\begin{align}
\bar{N}_E&=\bar{\omega} \bar{N} \label{relationlapse}\\
a_E&=\bar{\omega} a \label{relationscalefactor},
\end{align}
where we have decomposed the conformal factor $\omega$ in a background part plus a small (quantum) fluctuation
\begin{equation}
\omega(\Phi(x))=\bar{\omega}(t)+\delta\omega(t,\bvec{x}).\label{conformaldecomposition}
\end{equation}
 Now, in order to arrive at the Jordan frame Hamiltonian \eqref{backgroundhamiltonianshift2} from Eq. \eqref{backgroundhamiltonianEinstein} we see that the momenta, field derivative and the potential in the Einstein and Jordan frames are related as
\begin{align}
\mathcal{P}_E&=\frac{1}{\bar{\omega}}\mathcal{P} \label{relationEinsteinframemmomentumP}\\
\mathcal{P}_{\phi,E}&=\sqrt{\frac{\bar{\omega}^{D-2}}{\bar{\Omega}}}\hat{\mathcal{P}}_{\phi} \label{relationEinsteinframemmomentumPphi}\\
V_E(\phi_E)&=\frac{1}{\bar{\omega}^D}V(\phi_E(\phi))\label{relationpotential},
\end{align}
if we make the identification
\begin{equation}
F(\phi)= \bar{\omega} ^{D-2}.
\end{equation}
With these field redefinitions the Hamiltonian in the Jordan frame \eqref{backgroundhamiltonianshift2} can be derived from the Einstein frame Hamiltonian \eqref{backgroundhamiltonianEinstein}. Furthermore we can verify that $\mathcal{P}_E\partial_0 a_E\rightarrow \mathcal{P}\partial_0 a+\frac{1}{D-2}\frac{F'}{F}a\mathcal{P}\partial_0 \phi$ under the field redefinition of $\mathcal{P}_E$. What remains to be checked is the relation between $\partial_0\phi_E$ and $\partial_0\phi$. Since the canonical momentum $\hat{\mathcal{P}}_{\phi}$ depends on $\partial_0\phi$ in a specific way, we should find the expression for the canonical momentum $P_{\phi,E}$ in terms of $\partial_0\phi_E$. By varying the action \eqref{backgroundactionEinstein} with respect to $\mathcal{P}_E$ and $\mathcal{P}_{\phi,E}$ we can find the definition of the canonical momenta in the Einstein frame,
\begin{align}
\mathcal{P}_E&=-2(D-1)(D-2)a_E^{D-2}H_E\label{EinsteinframemmomentumP}\\
\mathcal{P}_{\phi,E}&=a_E^{D-1}\dot{\phi}_E\label{EinsteinframemmomentumPphi},
\end{align}
where $H_E=\dot{a}_E/a_E$ and $\dot{\phi}_E=\bar{N}_E^{-1}\partial_0 \phi_E$ is the dotted derivative in the Einstein frame. We now compare Eqs. \eqref{EinsteinframemmomentumP} and \eqref{EinsteinframemmomentumPphi} to the momenta in the Jordan frame \eqref{Ponshell} and \eqref{Pphionshell} and use the relations between Jordan and Einstein frame momenta in Eqs. \eqref{relationEinsteinframemmomentumP} and \eqref{relationEinsteinframemmomentumPphi}. This will give us the relation between the Hubble parameter and the background field in the Jordan and Einstein frames,
\begin{align}
H_E&=\frac{1}{\bar{\omega}}\left(H+\frac{\dot{\bar{\omega}}}{\bar{\omega}}\right)=\frac{1}{F^{\frac{1}{D-2}}}\left(H+\frac{1}{D-2}\frac{F'}{F}\dot{\phi}\right)\label{relationHubble}\\
\dot{\phi}_E&=\frac{1}{\bar{\omega}}\sqrt{\frac{\bar{\Omega}}{\bar{\omega}^{D-2}}}\dot\phi=\frac{1}{F^{\frac{1}{D-2}}}\sqrt{\frac{\bar{\Omega}}{F}}\dot\phi,
\label{relationdotphi}
\end{align}
where the dotted derivatives on the left- and right-hand sides are the reparametrization invariant dotted derivatives in the Einstein and Jordan frames, respectively\footnote{Note that Eq. \eqref{relationdotphi} corresponds to the nonlinear field redefinition which is commonly used in the Lagrange formulation to bring the kinetic terms in the Einstein frame into canonical form. See for example Ref. \cite{Hwang:1996np}, or Ref. \cite{Bezrukov:2007ep} for the specific nonminimal coupling term $\xi R \Phi^2$.}. Using the relation for $\dot{\phi}_E$ in Eq. \eqref{relationdotphi}, and the field redefinitions in Eqs. \eqref{relationlapse}-\eqref{relationpotential}, we finally find that we can derive the Jordan frame action \eqref{backgroundactionshift2} from the Einstein frame action \eqref{backgroundactionEinstein}. Since the background fields in the Jordan and Einstein frames are related by time-dependent rescalings, we thereby establish the physical equivalence between the two frames at the classical level both on- and off-shell. In the next section we will establish the equivalence of the Jordan and Einstein frame actions up to second order in (quantum) fluctuations.
{\section{\label{sec:freeaction} Free action for cosmological perturbations} }
In this section we will derive the free action for gauge invariant cosmological perturbations for all dynamical and constraint fields. A common approach is to fix a gauge by setting either scalar field or metric perturbations to zero, and then to solve for the lapse and shift perturbations from the linearized constraint equations, see for example Ref. \cite{Maldacena:2002vr} and Ref. \cite{Feng:2010ya} for minimal and nonminimal coupling, respectively. In this paper we do not solve any linearized constraint equations, nor do we use gauge freedom to set some fields to zero. Instead we keep all the fields up to second order in fluctuations \eqref{perturbedpij}-\eqref{perturbedN}. We find for the action \eqref{canonicalactionnonminimalshifted2} up to second order in perturbations
\begin{align}
\nonumber S^{(2)}=&\int d^Dx\biggl\{\frac{\mathcal{P}}{2(D-1)}\left(2\pi^{ij}h_{ij}\partial_0 a+a \pi^{ij}\partial_0 h_{ij}\right)+\hat{\mathcal{P}}_{\phi}\hat{\pi}_{\varphi}\partial_0 \varphi\\
\nonumber &+\frac{a\mathcal{P}}{(D-1)(D-2)}\biggl[\frac{F'}{F}\left(\pi^{ij}h_{ij}\partial_0\phi+\pi^{ij}\delta_{ij}\partial_0\varphi+h\partial_0 \varphi\right)+\frac12 (D-1)\left(\frac{F'}{F}\right)''\varphi^2\partial_0\phi\\
&+\left(\frac{F'}{F}\right)'\varphi\left((D-1)\partial_0\varphi+(\pi^{ij}\delta_{ij}+h)\partial_0\phi\right)\biggr]-\bar{N}\mathcal{H}^{(2)}-n\mathcal{H}^{(1)}-N_i\mathcal{H}^{i(1)}\biggr\},
\label{quadraticnonminimalaction}
\end{align}
where $h\equiv h^{ij}\delta_{ij}$. We refer the reader to Appendix A of Ref. \cite{Prokopec:2010be} for some intermediate steps in the derivation. In this quadratic action indices are raised and lowered by the Kronecker delta $\delta_{ij}$. The Hamiltonian up to first order in perturbations is
\begin{align}
\nonumber \mathcal{H}^{(1)}=&-a^{D-3}F\left[\partial_i\partial_{j}h^{ij}-\nabla^2 h\right]-\frac{1}{a^{D-3}}\frac{1}{F}\frac{\mathcal{P}^2}{2(D-1)^2(D-2)}\left[\pi^{ij}\delta_{ij}-\frac14 (D-5)h-\frac12 (D-1)\frac{F'}{F}\varphi\right]\\
&+\frac{1}{a^{D-1}}\frac{\hat{\mathcal{P}}_{\phi}^2}{2\bar{\Omega}}\left[2\hat{\pi}_{\varphi}-\frac12 h-\frac{\bar{\Omega}'}{\bar{\Omega}}\varphi\right]+a^{D-1}\left[\frac12 h V+V_{,\phi}\varphi\right] +2a^{D-3}F'\nabla^{2}\varphi,
\label{linearnonminimalhamiltonian}
\end{align}
where $\nabla^2=\partial_i \partial^{i}=\delta^{ij}\partial_i\partial_j$. The Hamiltonian up to second order in perturbations is
\begin{align}
\nonumber \mathcal{H}^{(2)}=&-Fa^{D-3}\left[-\frac14 h\nabla^2h+\frac12 h\partial^{i}\partial^{j}h_{ij}-\frac12 h_{ij}\partial^{i}\partial^{l}h_{jl}+\frac14 h^{ij}\nabla^2h_{ij}+\frac{F'}{F}\varphi(\partial^{i}\partial^{j}h_{ij}-\nabla^2h)\right]\\
\nonumber &+\frac{\mathcal{P}^2}{4(D-1)^2a^{D-3}F}\biggl[ \frac12 \pi^{ij}A_{ijkl}\pi^{kl}+\frac{\pi^{ij}}{D-2}(2(D-3)h_{ij}-h\delta_{ij}) +h_i^jh_j^{i}\\
\nonumber & -\frac{D-1}{D-2}\left(\frac14 h_i^jh_j^{i}+\frac18 h^2\right)
 -\frac{2F}{D-2}\left(\frac{1}{F}\right)'\varphi(\pi^{ij}\delta_{ij}+h-\frac14 (D-1)h)-\frac{F}{2}\frac{D-1}{D-2}\left(\frac{1}{F}\right)''\varphi^2\biggr]\\
\nonumber & +\frac{\hat{\mathcal{P}}_{\phi}^2}{2a^{D-1}\bar{\Omega}}\left[\hat{\pi}_{\varphi}^2+\frac14 h_i^jh_j^{i}+\frac18 h^2-h\hat{\pi}_{\varphi}+ \bar{\Omega}\left(\frac{1}{\bar{\Omega}}\right)'\varphi(2\hat{\pi}_{\varphi}-\frac12 h)+\frac{\bar{\Omega}}{2}\left(\frac{1}{\bar{\Omega}}\right)''\varphi^2 \right]\\
\nonumber &+a^{D-3}\frac12 \partial^{i}\varphi\partial_i\varphi+a^{D-1}\left[\left(-\frac14 h_i^jh_j^{i}+\frac18 h^2\right)V+\frac12 h \varphi V_{,\phi}+\frac12 V_{,\phi\phi}\varphi^2\right]\\
&+2\partial_i\left(a^{D-3}\left[\frac12 h F'\partial^{i} \varphi -F'h^{ij}\partial_j\varphi+F'' \varphi \partial^{i} \varphi\right]\right),
\label{quadraticnonminimalhamiltonian}
\end{align}
where $A_{ijkl}=\delta_{ik}\delta_{jl}+\delta_{il}\delta_{jk}-\frac{2}{D-2}\delta_{ij}\delta_{kl}$. Note that the final term in Eq. \eqref{quadraticnonminimalhamiltonian} is a total derivative term and vanishes, but we give this term explicitly for future purpose. Finally the momentum density up to first order in perturbations is
\begin{align}
\mathcal{H}^{i(1)}=\partial^{i}\varphi\frac{1}{a^2}\left(\hat{\mathcal{P}}_{\phi}+\frac{aP}{D-2}\frac{F'}{F}\right)-\frac{\mathcal{P}}{(D-1)a}\left(\partial_j\pi^{ij}+\partial_jh^{ij}-\frac12 \partial^{i}h\right).
\label{linearnonminimalNiconstraint}
\end{align}
Now that we have found the free action \eqref{quadraticnonminimalaction} we want to make a few remarks:
\begin{itemize}
\item The action \eqref{quadraticnonminimalaction} is quite complicated due to many coupled fields;
\item It is unclear what are the dynamical degrees of freedom in Eq. \eqref{quadraticnonminimalaction};
\item The action \eqref{quadraticnonminimalaction} is not explicitly gauge invariant.
\end{itemize}
Some clarification is in order. The action \eqref{quadraticnonminimalaction} contains many different fields (e.g. $\pi^{ij}$, $h^{ij}$, $\hat{\pi}_{\varphi}$, $\varphi$, $n$ and $N_i$) coupled in a nontrivial way. We know however that the $n$ and $N_i$ are non-dynamical and impose constraints on $h^{ij}$ and $\varphi$. Furthermore $n$ and $N_i$ are completely arbitrary and need to be fixed by imposing gauge conditions\cite{Prokopec:2010be}. The 14-dimensional phase space of $h^{ij}$ and $\varphi$ is therefore reduced to a $14-4-4=6$-dimensional physical phase space. Indeed, a well known result from cosmological perturbation theory is that there is only one dynamical scalar degree of freedom and two dynamical tensor degrees of freedom, corresponding to a 6-dimensional phase space. This is not at all obvious from the action \eqref{quadraticnonminimalaction}. Finally we remark that, being derived from a diffeomorphism invariant action \eqref{nonminimalaction}, the action \eqref{quadraticnonminimalaction} should be gauge invariant (\textit{i.e.} invariant under infinitesimal coordinate transformations). However it is difficult to see this from Eq. \eqref{quadraticnonminimalaction}. All fields transform in a specific way under a coordinate transformation, and it is only a special combination of the fields that will be gauge invariant.\\
In order to extract the three dynamical degrees of freedom and show the explicit gauge invariance of the action, we will only have to do one thing: decouple all fields by defining shifted fields that diagonalize the action. As it turns out, the shifted fields will all be gauge invariant and there will only be three dynamical degrees of freedom. As a bonus, the action acquires a nice and simple form. As a start it is convenient to use the scalar-vector-tensor decomposition of the spatial metric~\footnote{\label{footnote:notation}Our notation differs from the one used in most literature where $h=\text{Tr}(h_{ij})=2(D-1)\psi+2\nabla^2 E, h-\nabla^2\tilde{h}=2(D-1)\psi$, see \cite{Mukhanov:1990me}.}, see Ref. \cite{Prokopec:2010be},
\begin{align}
h_{ij}=\frac{\delta_{ij}}{D-1}h+\left(\partial_i\partial_j-\frac{\delta_{ij}}{D-1}\nabla^2\right)\tilde{h}+\partial_{\left(i\right.}h^T_{\left.j\right)}+h_{ij}^{TT},
\label{scalarvectortensordecomposition}
\end{align}
with
\begin{align}
\partial^{i}h_i^T=0,~~~~~~~~~~~~~\partial^{i}h_{ij}^{TT}=0=\partial^{j}h_{ij}^{TT}.
\end{align}
Furthermore we decompose the shift vector $N^{i}$ in its longitudinal and transverse components,
\begin{equation}
N_i=\partial_i S+N_{i}^T,~~~~~~~~~~~~~~~~\text{with}~~~~~~~~~~\partial^{i}N_i^T=0.
\label{shiftdecomposition}
\end{equation}
The action can now be diagonalized by defining shifted fields (see Appendix \ref{sec:appendixDiagonalisingaction} for a derivation and definitions of introduced variables)
\begin{align}
\hat{\pi}_{\varphi}&=\hat{\tilde{\pi}}_{\varphi}-\frac12 \hat{I}_{\varphi}\\
\pi^{ij}&=\tilde{\pi}^{ij}-\frac12\left(I_{ij}-\delta_{ij} I\right)\\
n&=\tilde{n}-\frac{\bar{N}}{2W}I_n\\
\nabla^2S&=\nabla^2\tilde{S}-\frac{1}{2(D-1)}\frac{1}{1-\alpha}\left[J-(D-2)\bar{N}a^2\nabla^2 (\dot{\tilde{h}}-J_{h_{ij}}\tilde{h})\right]\\
\partial_{\left(i\right.}N^T_{\left.j\right)}&=\partial_{\left(i\right.}\tilde{N}^T_{\left.j\right)}+\frac{a^2\bar{N}}{2}\left(\partial_{\left(i\right.}\dot{h}^T_{\left.j\right)} -J_{h_{ij}}\partial_{\left(i\right.}h^T_{\left.j\right)}\right),
\end{align}
and the Sasaki-Mukhanov variable~\footnote{From Footnote \ref{footnote:notation} it follows that Eq. \eqref{comovingCutvperturbation} yields the better known form $\tilde{\varphi}=\varphi-\frac{\dot{\phi}}{H}\psi$.}
\begin{equation}
\tilde{\varphi}=\varphi-
z_0(h-\nabla^2\tilde{h})\label{comovingCutvperturbation},
\end{equation}
where
\begin{equation}
z_0=\frac{\dot{\phi}}{2(D-1)H}.
\label{definitionz0}
\end{equation}
After tedious calculations, of which we present some intermediate results in Appendix \ref{sec:appendixDiagonalisingaction}, we obtain the free action
\begin{align}
\nonumber S^{(2)}&=\int d^{D-1}x \bar{N}dt a^{D-1}\Biggl\{\frac{z^2}{z_0^2}\left[\frac12\dot{\tilde{\varphi}}^2-\frac12 \left(\frac{\partial_i\tilde{\varphi}}{a}\right)^2+\frac12 \frac{z_0}{a^{D-1} z^2}\left[a^{D-1}\frac{z^2}{z_0^2}\dot{z_0}\right]^{\cdot}\tilde{\varphi}^2\right]+\frac{F}{4}\biggl[(\dot{h}_{ij}^{TT})^2-\Bigl(\frac{\partial h_{ij}^{TT}}{a}\Bigr)^2\biggr]\\
&-\frac{\hat{\mathcal{P}}_{\phi}^2}{2a^{2(D-1)}\bar{\Omega}}\hat{\tilde{\pi}}_{\varphi}^2-\frac{\mathcal{P}^2}{4(D-1)^2a^{2(D-2)}F}\tilde{\pi}^{ij} \frac{A_{ijkl}}{2}\tilde{\pi}^{kl}+\frac{W}{a^{D-1}\bar{N}^2}\tilde{n}^2+\frac{F}{a^4 \bar{N}^2}\left((1-\alpha)[\nabla^2 \tilde{S}]^2+[\partial_{\left(i\right.}\tilde{N}^T_{\left.j\right)}]^2\right)\Biggr\},
\label{freeAction}
\end{align}
where
\begin{equation}
z^2=\frac{1}{4(D-1)^2}\frac{\bar{\Omega} \dot{\phi}^2}{(H+\frac{1}{D-2}\frac{F'}{F}\dot{\phi})^2}.
\end{equation}
Note that by setting $F=\frac12 M_p^2\equiv 1$ that $z^2\rightarrow z_0^2$ and we obtain the well known result from gauge invariant cosmological perturbation theory for a minimally coupled scalar field (see also Ref. \cite{Prokopec:2010be}).\\
The action \eqref{freeAction} is our most important result. When we compare this new free action to the original free action from Eq. \eqref{quadraticnonminimalaction} we can make the following remarks:
\begin{itemize}
\item All shifted fields are decoupled in the action \eqref{freeAction};
\item The only dynamical degrees of freedom in \eqref{freeAction} are 1 scalar and 2 tensor degrees of freedom;
\item All shifted fields in Eq. \eqref{freeAction} are gauge-invariant up to linear order in coordinate transformations.
\end{itemize}
For a proof of the third point we refer the reader to Appendix \ref{sec:appendixgaugeinvariance}. Thus our tedious diagonalization procedure has paid off: we have obtained a simple, explicitly gauge invariant action with one propagating scalar field $\tilde{\varphi}$ and a propagating graviton $h^{TT}_{ij}$. A variation of the action with respect to the non-dynamical $\tilde{\pi}^{ij}$ and $\hat{\tilde{\pi}}_{\varphi}$ gives the linearized Hamilton equations of motion. On the other hand the variation with respect to $\tilde{n}$, $\tilde{S}$ and $\tilde{N}^{T}$ gives the solutions of the linearized constraint equations. Therefore the free action \eqref{freeAction} contains all the properties of linearized inflationary perturbations, as well as the transition between the Hamilton and Lagrange formalism.\\
In the gauge invariant form \eqref{freeAction} the scalar field $\tilde{\varphi}$ can be quantized and the true scalar propagator can be extracted from the action. If we would also know the gauge invariant cubic and quartic vertices (meaning we have to calculate the action up to fourth order in perturbations), we would be able to calculate quantum corrections to the inflaton potential. We would have to work much harder to make the action gauge invariant when we also include these higher order interaction terms. We leave this for future work. We emphasize that the action \eqref{freeAction} is gauge invariant up to linear order in coordinate transformations. If we include higher order terms the free action would still have the same form as Eq. \eqref{freeAction}, but the gauge invariant fields will now also consist of combinations of higher order field perturbations. This will affect, for example, canonical quantization. This fact makes the construction of a fully gauge invariant formalism a worthy effort. As a final comment we note that one can extend our treatment for models which contain non-canonical kinetic terms such as the DBI model \cite{Easson:2009wc}.

\subsection{\label{section:quantumequivalence} Quantum equivalence of Jordan and Einstein frames}
In section \ref{section:classicalequivalence} we showed the classical equivalence of the Jordan and Einstein frame actions in Hamiltonian form. Now we want to demonstrate the quantum equivalence of the Jordan and Einstein frames at the level of the free action. Let us first consider the dynamical scalar $\tilde{\varphi}$ in the Jordan frame free action Eq. \eqref{freeAction}. If we redefine the field $\tilde{\varphi}$ in terms of the comoving curvature perturbation in the Jordan frame $\mathcal{R}$
\begin{equation}
\mathcal{R}=-\frac{\tilde{\varphi}}{z_0}
\end{equation}
the scalar action becomes
\begin{equation}
S^{(2)}_{\mathcal{R}}=\int d^{D-1}x \bar{N}dt a^{D-1}z^2\left[\frac12\dot{\mathcal{R}}^2-\frac12 \left(\frac{\partial_i\mathcal{R}}{a}\right)^2\right].
\label{Jordanframeactioncomoving}
\end{equation}
On the other hand the form of the action for a minimally coupled scalar field (see Ref. \cite{Prokopec:2010be}) is
\begin{align}
\nonumber S^{(2)}_{\mathcal{R}_E}=&\int d^{D-1}x \bar{N}_E dt a_E^{D-1}z_E^2\left[\frac12\dot{\mathcal{R}}_E^2-\frac12 \left(\frac{\partial_i\mathcal{R}_E}{a_E}\right)^2\right],\\
z_E^2=&\frac{1}{4(D-1)^2}\frac{\dot{\phi}_E^2}{H_E^2},
\label{Einsteinframeactioncomoving}
\end{align}
where the dotted derivative here means $\dot{\mathcal{R}}_E=\bar{N}_E^{-1}\partial_0\mathcal{R}_E$ and $\mathcal{R}_E$ is the comoving curvature perturbation in the Einstein frame. The prefactor in the action \eqref{Einsteinframeactioncomoving} only depends on the background fields and can therefore be transformed to a physically equivalent prefactor by performing a conformal transformation. Indeed, we can derive the action in the Jordan frame \eqref{Jordanframeactioncomoving} from the action in the Einstein frame \eqref{Einsteinframeactioncomoving} by a redefinition of the background fields as in Eqs. \eqref{relationlapse}, \eqref{relationscalefactor}, \eqref{relationHubble} and \eqref{relationdotphi}. The free actions \eqref{Jordanframeactioncomoving} and \eqref{Einsteinframeactioncomoving} are however only truly physically equivalent if the comoving curvature perturbation does not change under a conformal transformation, \textit{\textit{i.e.}} $\mathcal{R}=\mathcal{R}_E$. This can be proved in the following way. If we decompose the conformal factor and the metric in a background plus a perturbed part as in Eqs. \eqref{conformaldecomposition} and \eqref{perturbedgij} we can show that (see also Appendix \ref{sec:appendixconformalinvariance})
\begin{equation}
(h_E-\nabla^2\tilde{h}_E)=(h-\nabla^2\tilde{h})+2(D-1)\frac{\delta\omega}{\bar{\omega}}.
\end{equation}
Now we want to know how the scalar inflaton fluctuation in the Einstein frame $\varphi_E$ is related to $\varphi$ in the Jordan frame. Suppose now we have a scalar field $\Phi_E$ in the Einstein frame action which is a function of the scalar field $\Phi$ in the Jordan frame. We also wish to decompose this $\Phi_E$ into a background part plus a small quantum fluctuation by substituting $\Phi=\phi+\varphi$. This gives
\begin{equation}
\Phi_E(\phi+\varphi)=\Phi_E(\phi)+\frac{\partial \Phi_E}{\partial\phi}\varphi=\Phi_E(\phi)+\frac{\partial_0\Phi_E}{\partial_0 \phi}\varphi\equiv \phi_E+\varphi_E.
\end{equation}
This leads to the convenient relation at the level of linearized perturbations
\begin{equation}
\frac{\varphi_E}{\dot{\phi}_E}=\bar{\omega}\frac{\varphi}{\dot{\phi}}=\bar{\omega}\frac{\delta\omega}{\dot{\bar{\omega}}}\label{relationvarphi},
\end{equation}
where the extra factor of $\bar{\omega}$ appears because of the difference between the dotted derivatives in the Einstein and the Jordan frame. The last relation is true since the conformal transformation is a function of $\Phi$, \textit{i.e.} $\omega=\omega(\Phi)$. With this relation and the Hubble parameter in the Einstein frame from Eq. \eqref{relationHubble}, we find
\begin{equation}
\frac{H_E}{\dot{\phi}_E}\varphi_E=\frac{H}{\dot{\phi}}\varphi+\frac{\delta\omega}{\bar{\omega}},
\end{equation}
such that
\begin{equation}
\mathcal{R}_E=(h_E-\nabla^2\tilde{h}_E)-2(D-1)\frac{H_E}{\dot{\phi}_E}\varphi_E =(h-\nabla^2\tilde{h})-2(D-1)\frac{H}{\dot{\phi}}\varphi=\mathcal{R}.
\end{equation}
Thus the comoving curvature perturbation is invariant under a conformal transformation up to linear order in perturbations. This was first proved by Makino and Sasaki\cite{Makino:1991sg} and Fakir et al. \cite{Fakir:1992cg}. In fact one can show that the comoving curvature perturbation is conformally invariant in the fully nonlinear approach, see Ref. \cite{Chiba:2008ia}. Therefore we have established the equivalence of the Jordan and Einstein frame scalar actions at the classical level as well as at the level of quadratic (quantum) fluctuations.\\
Now that we have checked the physical equivalence for the scalar sector, let us see how the rest of the action \eqref{freeAction} transforms under the conformal transformation. First of all, let us give the Einstein frame action for the graviton and constraint fields,
\begin{align}
\nonumber S^{(2)}=\int d^{D-1}x \bar{N}_Edt a_E^{D-1}\Biggl\{&\frac{1}{4}\biggl[(\dot{h}_{ij,E}^{TT})^2-\Bigl(\frac{\partial h_{ij,E}^{TT}}{a_E}\Bigr)^2\biggr]-\frac{\hat{\mathcal{P}}_{\phi,E}^2}{2a_E^{2(D-1)}}\tilde{\pi}_{\varphi,E}^2-\frac{\mathcal{P}_E^2}{4(D-1)^2a_E^{2(D-2)}}\tilde{\pi}_E^{ij} \frac{A_{ijkl}}{2}\tilde{\pi}_E^{kl}\\
&+\frac{W_E}{a_E^{D-1}\bar{N}_E^2}\tilde{n}_E^2+\frac{1}{a_E^4 \bar{N}_E^2}\left((1-\alpha_E)[\nabla^2 \tilde{S}_E]^2+[\partial_{\left(i\right.}\tilde{N}^T_{\left.j\right),E}]^2\right)\Biggr\},
\label{freeActionEinstein}
\end{align}
where $W_E$ and $\alpha_E$ are defined in Eqs. \eqref{definitionW} and \eqref{definitionalpha} of Appendix \ref{sec:appendixDiagonalisingaction}, where the subscript $E$ denotes that these quantities depend on the Einstein frame background fields. Now we perform the conformal transformation of the action \eqref{freeActionEinstein} using Eqs. \eqref{relationlapse}, \eqref{relationscalefactor} and \eqref{relationEinsteinframemmomentumP} - \eqref{relationpotential}. We find that the Einstein frame action transforms to the Jordan frame action \eqref{freeAction} if the graviton and constraint fields transform as
\begin{align}
\tilde{n}_E&=\bar{\omega}\tilde{n}\label{conformaltransfn}\\
\tilde{S}_E&=\bar{\omega}^2 \tilde{S}\\
\tilde{N}^T_{i,E}&=\bar{\omega}^2\tilde{N}^T_{i}\label{conformaltransformNiT}\\
h_{ij,E}^{TT}&=h_{ij}^{TT}\label{conformaltransfhij}\\
\tilde{\pi}_{\varphi,E}&=\hat{\tilde{\pi}}_{\varphi}\\
\tilde{\pi}_E^{ij}&=\tilde{\pi}^{ij}\label{conformaltransfpiij}.
\end{align}
In Appendix \ref{sec:appendixconformalinvariance} we show that the graviton and constraint fields transform in precisely this way under a conformal transformation\footnote{The gauge invariant lapse perturbation $\tilde{n}$ can be made invariant under a conformal transformation if we define the lapse perturbation as $N=\bar{N}(t)(1+n(t,\bvec{x}))$ as compared to Eq. \eqref{perturbedN}. Moreover, the lapse part of the free action would be time reparametrization invariant. Furthermore we could have defined the shift perturbations as $N_i=a\bar{N}n_i$, where the diagonalized shift perturbation $\tilde{n}_i$ would also be invariant under a conformal transformation and the action as a whole would be time reparametrization invariant. Note also that, with these definitions of the perturbed fields, every spatial derivative in the action \eqref{freeAction} as well as \eqref{quadraticnonminimalaction} contains a factor $a^{-1}$.}. Therefore the complete free Jordan frame \eqref{freeAction} and Einstein frame \eqref{Einsteinframeactioncomoving}+\eqref{freeActionEinstein} actions are physically equivalent.

{\section{Higgs inflation}\label{sec:Higgsinflation} }
Recently Bezrukov and Shaposhnikov\cite{Bezrukov:2007ep} revived the old idea by Salopek, Bond and Bardeen \cite{Salopek:1988qh} that the Higgs boson can be the inflaton field if it is nonminimally coupled to gravity. The requirement for Higgs inflation is a large nonminimal coupling $|\xi|\gg 1$, ensuring the flatness of the Higgs potential for large field values. Since then there has been much debate whether or not quantum corrections destroy the flatness of the Higgs potential, thereby spoiling Higgs inflation. One and two loop corrections have been calculated in both the Einstein\cite{Bezrukov:2008ej,DeSimone:2008ei,Bezrukov:2009db} and Jordan\cite{Barvinsky:2008ia,Barvinsky:2009fy,Barvinsky:2009ii} frames. Although there is some debate about the calculational methods, all loop calculations predict that Higgs inflation is valid if the Higgs mass lies in a specific range, testable by the LHC.\\
In Refs. \cite{Bezrukov:2008ej,DeSimone:2008ei,Bezrukov:2009db,Barvinsky:2008ia,Barvinsky:2009fy,Barvinsky:2009ii} the validity of Higgs inflation was tested in the inflationary regime. Here the Higgs boson has a large expectation value $\langle H \rangle\geq M_P/\sqrt{|\xi|}$ and slowly rolls down the inflaton potential. Recently \cite{Burgess:2009ea,Lerner:2009na,Hertzberg:2010dc,Burgess:2010zq} however, the validity of Higgs inflation was questioned in the small-field limit where the Higgs expectation value is $\langle H \rangle=v=246 \text{GeV}$. Hertzberg\cite{Hertzberg:2010dc} considered the general case of a theory with multiple scalar fields. It was found that, for the pure gravity and kinetic sectors, the small field effective theory has a cut-off at an energy scale of $M_P$ if there is only one scalar field, but when more than one scalar field is involved the cut-off is $M_P/|\xi|$. In the Jordan frame the cut-off $M_P/|\xi|$ can be almost directly read off from the scalar-graviton interaction term. However, when considering scalar-scalar scattering via graviton exchange, the lowest order diagrams add up to zero for a single scalar field. Therefore the actual cut-off scale is $M_P$. In the Einstein frame this is even more clear. After the conformal transformation the cut-off appears in a dimension 6 scalar kinetic term, but this term can be removed via a nonlinear field redefinition.\\
In the case of multiple scalar fields the above reasoning no longer applies. In the Jordan frame the lowest order diagrams do not vanish because the scalar fields are not identical, giving the cut-off $M_P/|\xi|$. In the Einstein frame the unitarity violating kinetic term cannot be removed by a field redefinition, because it is in general not possible to bring the kinetic term into canonical form for multiple scalar fields (see Ref. \cite{Kaiser:2010ps} for more details). The arguments above apply to the pure gravity and kinetic sectors of the theory, but even for the single field case Hertzberg \cite{Hertzberg:2010dc} finds that scalar self-interactions due to the non-polynomial potential in the Einstein frame most likely cause unitarity problems at the scale $M_P/|\xi|$.
\\
Now we switch to the Standard Model. In this case the Higgs doublet contains in principle 4 scalar fields, but the 3 Goldstone bosons are eaten up by the $W^{\pm}$ and $Z$ bosons. Therefore one might wonder if the cut-off shows up in the terms containing these gauge bosons. Indeed, Burgess, Lee and Trott \cite{Burgess:2010zq} showed that in the Standard Model the cut-off scale $M_P/|\xi|$ appears in the Higgs-gauge interactions.\\
Now the crucial point is that the cut-off $M_P/|\xi|$ of the small-field effective theory is very close to the energy scale at the end of inflation $H_{\text{end}}\simeq \sqrt{\lambda/12} M_P/|\xi|$ (where $0.11<\lambda\lesssim 0.27$ at the electroweak scale), which is also the point where the small-field limit becomes valid. This means that higher order operators, needed to solve the unitarity problems at the cut-off scale $M_P/|\xi|$ in the small-field effective theory, will affect the inflationary theory and thereby destroy Higgs inflation. Therefore it seems that Higgs inflation is ruled out as a valid theory.\\
In contrast to the previous arguments, Bezrukov et al. \cite{Bezrukov:2010jz} very recently showed that the effective cut-off actually depends on the expectation value of the Higgs inflaton field. An intermediate region was identified for field values $M_p/|\xi|<\langle \phi \rangle < M_P/\sqrt{|\xi|}$ where the cut-off scale scales as $\Lambda = |\xi|\langle\phi\rangle^2/M_P$. The authors showed that all relevant energy scales throughout the evolution of the universe are below the corresponding cut-off scale. Still quantum corrections could spoil the unitarity of Higgs inflation, and a systematic way of obtaining quantum loop corrections has been proposed.\\
Considering the ongoing discussion about the unitarity of Higgs inflation, we would like to make a few remarks. First of all there are so far no rigorous calculations of quantum corrections to the Higgs potential or Higgs-gauge interactions in the small field limit ($\langle \phi \rangle < M_P/|\xi|$) or the intermediate region ($M_p/|\xi|<\langle \phi \rangle < M_P/\sqrt{|\xi|}$). Secondly, the cut-off scale is found in the Jordan frame by considering Higgs-graviton interactions. As we have shown before, the inflaton perturbation actually combines with the scalar part of the metric to form one 'gauge' invariant variable. Therefore, in order to consistently calculate quantum corrections to either the Higgs potential or Higgs-gauge interactions, we need to construct the completely diffeomorphism\footnote{We use the terminology "diffeomorphism invariance" here instead of the previously used "gauge invariance" in order to avoid confusion with the well known concept of gauge freedom in the Standard Model} invariant Higgs action. In the previous section we derived the free action for a single inflaton field. In this section we apply this to the Standard Model Higgs action with a nonminimal coupling to gravity. The action reads
\begin{equation}
S=\int d^D x \sqrt{-g}\left\{-\left(\frac{M_P^2}{2}-\xi H^{\dagger} H\right)R-g^{\mu\nu}(D_{\mu} H)^{\dagger}D_{\nu} H -\lambda \left(H^{\dagger} H-\frac{v^2}{2}\right)^2\right\},
\label{Higgsaction}
\end{equation}
where $H$ is the complex Higgs doublet with vev $\langle H \rangle_0=v/\sqrt{2}$ and
\begin{equation}
D_{\mu}H=\left(\partial_{\mu}-igA_{\mu}^{a}\tau^{a}-i\frac12 g' B_{\mu}\right)H,
\end{equation}
is the covariant derivative with $A_{\mu}^{a}$ and $B_{\mu}$ the $SU(2)$ and $U(1)$ gauge bosons with coupling constants $g$ and $g'$, and $\tau^{a}=\sigma^{a}/2$.\\
Now, in conventional chaotic inflationary scenarios the inflaton field is a real scalar field with a large classical expectation value. The Higgs doublet in Eq. \eqref{Higgsaction} contains two complex scalar fields, thus it is not clear what the inflaton field is. This becomes more obvious when we choose the following decomposition of the Higgs doublet
\begin{equation}
H=\frac{\Phi}{\sqrt{2}}\exp{({i\tau^{a} \alpha^{a}})}\cdot \left(\begin{array}{c} 0\\ 1 \end{array}\right),
\label{Higgsdecomposition}
\end{equation}
where $\Phi$ and the $\alpha^{a}$ are now four real scalar fields and the projection vector $(1,0)^T$ ensures that $H$ is a doublet. In this decomposition it is easy to see that $H^{\dagger} H=\frac12 \Phi^2$. Furthermore, by a redefinition $\tilde{A}_{\mu}^{a}=A_{\mu}^{a}-\frac{1}{g}\partial_{\mu}\alpha^{a}-i\alpha^{a}A_{\mu}^{b}[\tau^{a},\tau^{b}]$ we can absorb the three would-be Goldstone bosons $\alpha^{a}$ into the gauge bosons $A_{\mu}^{a}$. In fact, we can always perform an $SU(2)$ rotation on the Higgs doublet in Eq. \eqref{Higgsdecomposition} such that the three would-be Goldstone bosons disappear, which corresponds to fixing the unitary gauge. If we now define
\begin{align}
W^{\pm}_{\mu}&=\frac{1}{\sqrt{2}}\left(A_{\mu}^{1}\mp i A_{\mu}^{2}\right)\\
Z_{\mu}^{0}&=\frac{1}{\sqrt{g^2+g^{\prime 2}}}\left(g A_{\mu}^{3}-g'B_{\mu}\right),
\end{align}
the action \eqref{Higgsaction} becomes
\begin{align}
S=\int d^D x \sqrt{g}\Biggl\{& -\frac12(M_P^2-\xi \Phi^2)R-\frac12 g^{\mu\nu}\partial_{\mu} \Phi\partial_{\nu}\Phi -\frac14\lambda (\Phi^2-v^2)^2\\
&-\frac{m_W^2}{v^2}g^{\mu\nu}W_{\mu}^{+}W_{\nu}^{-}\Phi^2+\frac12 \frac{m_Z^2}{v^2} g^{\mu\nu}Z^0_{\mu}Z^0_{\nu}\Phi^2\Biggr\},
\label{Higgsactionunitary}
\end{align}
with $m_W^2=\frac14 g^2 v^2$ and $m_Z^2=\frac14 (g^2+g^{\prime 2})v^2$. We see that the first part of the action \eqref{Higgsactionunitary} is equal to the action \eqref{nonminimalaction} with the identification $F(\Phi)=\frac12(M_P^2-\xi \Phi^2)$ and $V(\Phi)=\frac14\lambda (\Phi^2-v^2)^2$. The second part contains the Higgs-gauge interaction terms. If we now want to calculate the free gauge invariant action for the Higgs sector of the SM we can simply do the expansions \eqref{perturbedpij}-\eqref{perturbedN} with $\Phi=\phi(t)+\varphi(t,\bvec{x})$, which will result in \eqref{freeAction}. The field $\varphi$ is in this case identified with the Higgs boson and $\phi$ is the classical background field with vev $v$. Since the gauge bosons $W^{\pm}_{\mu}$ and $Z^{0}_{\mu}$ are pure fluctuations, the free action \eqref{freeAction} will have additional terms
\begin{equation}
S^{(2)}_{\text{add}}=\int d^{D-1}x dt a^{D-1}\Biggl\{-\frac{m_W^2}{v^2}\bar{g}^{\mu\nu}W_{\mu}^{+}W_{\nu}^{-}\phi^2+\frac12 \frac{m_Z^2}{v^2} \bar{g}^{\mu\nu}Z^0_{\mu}Z^0_{\nu}\phi^2\Biggr\},
\end{equation}
where $\bar{g}^{\mu\nu}=\text{diag}(-\bar{N}^{-2},a^{-2}\delta^{ij})$ is the background ADM metric. So in the end we have shown that the free Standard Model with a nonminimally coupled Higgs boson, which has a local $SU(2)$ symmetry, can be written in terms of one diffeomorphism invariant scalar $\tilde{\varphi}$ \eqref{comovingCutvperturbation} and three mass terms for the gauge bosons. The free action can be used to extract the diffeomorphism invariant propagators for the Higgs inflaton field and gauge bosons. If we want to calculate quantum corrections to the free propagators and the Higgs potential in an invariant manner, we need to find the gauge invariant action up to third and fourth order in perturbations. We will leave this for future work. The analysis in this section shows that, when the backreaction from the $W^{\pm}$ and $Z$ bosons is neglected, the single scalar field and $SU(2)$ Higgs doublet lead to identical quadratic actions for cosmological perturbations in nonminimally coupled models.\\
The approach that we used in this section can be applied in a much more general setting than the Standard Model. In fact, any theory with a local $SU(N)$ or $SO(N)$ symmetry (for example GUT theories) can be written in terms of one dynamical scalar and a number of massive gauge bosons if we use the suitable generalization of \eqref{Higgsdecomposition}. Therefore it is always possible in these theories to have a single light inflaton field, thus opening the way for an inflationary era.
\bigskip
\bigskip

{\section{Discussion}\label{sec:discussion}}
In this paper we derived the free perturbation action for the nonminimally coupled inflaton field in the Jordan frame. By working in the ADM formalism we could split the metric tensor into scalar, vector, tensor and constraint fields. By performing a diagonalization procedure we obtained the main result \eqref{freeAction}. In this form the action is explicitly gauge invariant up to linear order in coordinate transformations and the only propagating degrees of freedom are the scalar inflaton field and the graviton. We showed that the Jordan frame action can be derived from the Einstein frame action by performing a conformal transformation of the metric, thereby establishing the physical equivalence of the two frames at the level of quadratic fluctuations.\\
In order to calculate quantum corrections to the inflaton potential we need an unambiguous, gauge invariant action. So far we have obtained the free gauge invariant action, what remains is the higher order gauge invariant action, with which one can perform definite calculations that will tell us whether or not nonminimal inflation is spoiled by quantum corrections. The one loop effective inflaton potential was already found in de Sitter space in the limit where $|\xi|\ll 1$, see Ref. \cite{Bilandzic:2007nb}; what remains to be done is to calculate $V_{eff}$ in the limit $|\xi|\gg 1$ for more general cosmological backgrounds \cite{Janssen:2009pb,Sloth:2006az,Seery:2007wf,Burgess:2009bs}. These thorough calculations are especially needed in the context of the Higgs inflation model, which at the moment seems to possess troublesome large quantum corrections to the small field effective theory. We hope to give a definite conclusion concerning this question in future work.
\bigskip
\bigskip

\section*{Acknowledgements}
We would like to thank Sander Mooij, Marieke Postma and Damien George for useful discussions, and especially for their significant contribution to section \ref{sec:Higgsinflation}. This research was supported by the Dutch Foundation for 'Fundamenteel Onderzoek der Materie' (FOM) under the program "Theoretical particle physics in the era of the LHC", program number FP 104.

\appendix
\section{\label{secappendixADMaction}Canonical form of the action for a nonminimally coupled scalar field}

In this section we will show how to obtain Eq. \eqref{canonicalactionnonminimal} from Eq. \eqref{nonminimalaction}. We start by substituting the ADM metric \eqref{ADMlineelement} into Eq. \eqref{nonminimalaction} and we find the standard ADM action (see Ref. \cite{Arnowitt:1962hi})
\begin{align}
\nonumber S=&\int d^Dx \sqrt{g}\Biggl\{\biggl[-g^{ij}\partial_0 K_{ij}-\partial_0 K+N(R+K^2-K_{ij}K^{ij})-2N_i\nabla_j(K^{ij}-g^{ij}K)\\
&-2g^{ij}\nabla_j(\nabla_iN-N^{k}K_{ik})\biggr]F(\Phi)+\frac{1}{2}\frac{1}{N}\left(\partial_0{\Phi}-N^{i}\partial_i\Phi\right)^2-N\frac{1}{2}g^{ij}\partial_i\Phi\partial_j\Phi-NV(\Phi)\Biggr\},
\label{ADMactionnonminimal}
\end{align}
where $K_{ij}=-N{^D\Gamma^{0}_{ij}}=-\frac{1}{2N}\left(\partial_0g_{ij}-\nabla_iN_j-\nabla_jN_i\right)$ and $K=g^{ij}K_{ij}$. It is understood that the only dynamical metric field present in the action is the spatial metric $g_{ij}$, such that the covariant derivatives as well as $R$ only depend on this metric. Now we want to find the canonical momentum $p^{ij}$ and $p_{\Phi}$ conjugate to $g_{ij}$ and $\Phi$ respectively. In order to find $p^{ij}$ we have to vary the action with respect to $\partial_0g_{ij}$, which is up to a factor $2N$ equal to varying the action with respect to $K_{ij}$. Since the action \eqref{ADMactionnonminimal} contains derivatives of $K_{ij}$ we remove these by performing partial integrations. The action then takes the simpler form
\begin{align}
\nonumber S=&\int d^Dx \sqrt{g}\Biggl\{N\left(R+K^{ij}K_{ij}-K^2\right)F+2KF'\left(\partial_0{\phi}- N^{i}\partial_i\phi\right)-2N g^{ij}\nabla_i\nabla_jF(\phi)\\
&+\frac{1}{2}\frac{1}{N}\left(\partial_0{\Phi}-N^{i}\partial_i\Phi\right)^2-N\frac{1}{2}g^{ij}\partial_i\Phi\partial_j\Phi-NV(\Phi)\Biggr\}.
\label{ADMactionnonminimalafterpartialintegration}
\end{align}
The canonical momenta can now easily be extracted and are
\begin{align}
p^{ij}\equiv \frac{\delta S}{\delta \partial_0 g_{ij}}&=\sqrt{g}\left(Kg^{ij}-K^{ij}\right)F-\frac{1}{N}\sqrt{g}g^{ij}F'\left(\partial_0{\Phi}- N^{i}\partial_i\Phi\right)&
\label{nonminimalmomentum}\\
p_{\Phi}\equiv \frac{\delta S}{\delta \partial_0\Phi}&=\sqrt{g}\frac{1}{N}\left(\partial_0{\Phi}-N^{i}\partial_i\Phi\right)+2\sqrt{g} K F'.&
\label{nonminimalscalarmomentum}
\end{align}
We now want to write our action in the form of Eq. \eqref{canonicalactionnonminimal} by substituting the momenta into Eq. \eqref{ADMactionnonminimalafterpartialintegration}. First we have to express $K$ and $K_{ij}$ in terms of the momenta, which gives
\begin{align}
K&=\frac{1}{\sqrt{g}F}\frac{1}{D-2}\frac{p+(D-1)F'p_{\Phi}}{\Omega}&\label{Kintermsofpi}\\
K^{ij}&=\frac{1}{\sqrt{g}F}\left[\frac{1}{D-2}g^{ij}\frac{1+2\frac{F^{\prime 2}}{F}}{\Omega}p-p^{ij}+\frac{1}{D-2}g^{ij}\frac{F'}{\Omega}p_{\Phi}\right],
\label{Kijintermsofpi}
\end{align}
where $p=p^{ij}g_{ij}$ and $\Omega$ is given in Eq. \eqref{definitionOmega}. The canonical form of the action \eqref{ADMactionnonminimal} is
\begin{equation}
S=\int d^D x \left[p^{ij}\partial_0 g_{ij}+p_{\phi}\partial_0{\phi}-H(p^{ij},g^{ij},p_{\Phi},\Phi)\right],
\label{ADMcanonicalactionnonminimal}
\end{equation}
where the Hamiltonian is
\begin{equation}
H=p^{ij}\partial_0 g_{ij}+p_{\phi}\partial_0{\phi}-\sqrt{g}\mathcal{L},
\end{equation}
with $\mathcal{L}$ the Lagrangian density from the action \eqref{ADMactionnonminimal} (or the action \eqref{ADMactionnonminimalafterpartialintegration} after partial integrations). The final canonical form of the action in Eq. \eqref{canonicalactionnonminimal} is obtained by substituting Eqs. \eqref{Kintermsofpi} and \eqref{Kijintermsofpi} into this Lagrangian density, thereby expressing the Hamiltonian in terms of the canonical momenta.
{\section{Diagonalizing the quadratic action}\label{sec:appendixDiagonalisingaction}}
\subsubsection*{Momenta sector}
We first consider the sector of the action \eqref{quadraticnonminimalaction} containing only the momenta $\pi^{ij}$ and $\hat{\pi}_{\varphi}$. We can write this as
\begin{equation}
S_{\pi}^{(2)}=\int d^Dx\bar{N}\left\{-\frac{\hat{\mathcal{P}}_{\phi}^2}{2a^{D-1}\bar{\Omega}}\left[\hat{\pi}_{\varphi}^2+\hat{I}_{\varphi}\hat{\pi}_{\varphi}\right] -\frac{\mathcal{P}^2}{4(D-1)^2a^{D-3}F}\left[\frac12 \pi^{ij}A_{ijkl}\pi^{kl}+I_{ij}\pi^{ij}\right]\right\},
\label{quadraticactionmomentaonly}
\end{equation}
where $A_{ijkl}=\delta_{ik}\delta_{jl}+\delta_{il}\delta_{jk}-\frac{2}{D-2}\delta_{ij}\delta_{kl}$ and
\begin{align}
\hat{I}_{\varphi}=&-h -\frac{2 a^{D-1}\bar{\Omega}}{\hat{\mathcal{P}}_{\phi}}\dot{\varphi}+\frac{2}{\bar{N}}n+2\bar{\Omega}\left(\frac{1}{\bar{\Omega}}\right)'\varphi
\label{shigtinpiphi}\\
\nonumber I_{ij}=&-\frac{2(D-1)}{\mathcal{P}}a^{D-2}F\dot{h_{ij}}+\frac{2}{D-2}\left[(D-3)-\frac{2(D-1)(D-2)a^{D-2}F}{\mathcal{P}}\left(H+\frac{1}{D-2}\frac{F'}{F}\dot{\phi}\right)\right]h_{ij}\\
\nonumber &-\frac{\delta_{ij}}{D-2}h-\frac{\delta_{ij}}{\bar{N}}\frac{2}{D-2}n+\frac{4(D-1)}{\bar{N}\mathcal{P}}Fa^{D-4}\partial_{\left(i\right.}N_{\left.j\right)}\\
& -\left[\frac{4(D-1)a^{D-2}}{(D-2)\mathcal{P}}\left(F'\dot{\varphi}+F\left(\frac{F'}{F}\right)'\varphi\dot{\phi}\right)+\frac{2}{D-2}F\left(\frac{1}{F}\right)'\varphi\right]\delta_{ij}.
\label{shiftinpiij}
\end{align}
Note that on-shell, \textit{i.e.} using Eq. \eqref{Ponshell}, the term multiplying $h_{ij}$ in Eq. \eqref{shiftinpiij} simplifies to $2 h_{ij}$. If we now define shifted momenta
\begin{align}
\hat{\tilde{\pi}}_{\varphi}=&\hat{\pi}_{\varphi}+\frac12 \hat{I}_{\varphi} \label{rhophi}\\
\tilde{\pi}^{ij}=&\pi^{ij}+\frac12\left(I_{ij}-\delta_{ij} I\right)\label{rhopiij},
\end{align}
where $I\equiv \delta^{ij} I_{ij}$, we find that we can write the momentum action in diagonalized form,
\begin{equation}
S_{\pi}^{(2)}=\int d^Dx\bar{N}\left\{-\frac{\hat{\mathcal{P}}_{\phi}^2}{2a^{D-1}\bar{\Omega}}\left[\hat{\tilde{\pi}}_{\varphi}^2-\frac14 \hat{I}_{\varphi}^2\right] -\frac{\mathcal{P}^2}{4(D-1)^2a^{D-3}F} \left[\tilde{\pi}^{ij}\frac{A_{ijkl}}{2}\tilde{\pi}^{kl}-\frac14 \left(I_{ij}I^{ij}-I^2\right)\right]\right\}.
\label{quadraticactionmomentaonlydiagonalised}
\end{equation}
The extra terms containing $\hat{I}_{\varphi}^2$ and $\left(I_{ij}I^{ij}-I^2\right)$ now give quadratic contributions of $n^2$ and $|\partial_{\left(i\right.}N_{\left.j\right)}|^2$, which allows us to diagonalize also the terms containing these factors.\\
\subsubsection*{Lapse sector}
From the diagonalized momentum sector of the action, see \eqref{quadraticactionmomentaonlydiagonalised}, we collect all the terms containing the lapse function $n$. We collect these terms from the rest of the action \eqref{quadraticnonminimalaction} as well, which are only in the $-n \mathcal{H}^{(1)}$ term. We find that we can write
\begin{equation}
S_n^{(2)}=\int d^Dx \bar{N}\left\{W \frac{n^2}{\bar{N}^2}+ \frac{I_n n}{\bar{N}}\right\},
\label{quadraticactionlapseonly}
\end{equation}
where
\begin{align}
W=&\frac{1}{F}\frac{1}{a^{D-3}}\frac{-\mathcal{P}^2}{4(D-1)(D-2)}+\frac{\hat{\mathcal{P}}_{\phi}^2}{2\bar{\Omega} a^{D-1}}\xrightarrow{\text{on-shell}}-a^{D-1}V\label{definitionW}\\
\nonumber I_n=&-\frac{a\mathcal{P}}{2(D-1)}\dot{h}+a^{D-3}F\left[\partial_i\partial_{j}h^{ij}-\partial_i\partial^{i} h\right]-\left(\hat{\mathcal{P}}_{\phi}+\frac{a\mathcal{P}F'}{(D-2)F}\right)\dot{\varphi}-2a^{D-3}F'\partial_i\partial^{i}\varphi\\
&+C_{\varphi}\varphi-C_h\frac{h}{2}+\frac{\mathcal{P}}{(D-1)a\bar{N}}\partial^{i}N_i
\label{shiftinn}
\end{align}
with
\begin{align}
\nonumber C_{\varphi}&=-a^{D-1}V_{,\phi}+\frac{1}{F}\frac{1}{a^{D-3}}\frac{\mathcal{P}^2}{4(D-1)(D-2)}\frac{F'}{F}-\frac{\hat{\mathcal{P}}_{\phi}^2}{2\bar{\Omega} a^{D-1}}\frac{\bar{\Omega}'}{\bar{\Omega}}-\frac{a\mathcal{P}}{(D-2)}\left(\frac{F'}{F}\right)'\dot{\phi}\\
&\xrightarrow{\text{on-shell}}a^{D-1}\left[\ddot{\phi}+(1+2F'')(D-1)H\dot{\phi}-2(D-1)F'(H^2+\dot{H})\right]\label{nshiftvarphi}\\
\nonumber C_h&=\frac{-\mathcal{P}^2}{4(D-1)(D-2)a^{D-3}F}\left[\frac{D-5}{D-1}-\frac{8(D-2)a^{D-2}F}{\mathcal{P}}\left(H+\frac{1}{D-2}\frac{F'}{F}\dot{\phi}\right)\right] +\frac{\hat{\mathcal{P}}_{\phi}^2}{2\bar{\Omega} a^{D-1}}+a^{D-1}V \label{nshifth}\\
&\xrightarrow{\text{on-shell}}0.
\end{align}
Note that on-shell, these expressions simplify greatly. We now shift the lapse,
\begin{equation}
\tilde{n}=n+\frac{\bar{N}}{2W}I_n,
\label{tilden}
\end{equation}
such that we can write
\begin{equation}
S_n^{(2)}=\int d^Dx \bar{N}\left\{W \frac{\tilde{n}^2}{\bar{N}^2}- \frac14 \frac{1}{W}I_n^2\right\}.
\label{quadraticactionlapsediagonalised}
\end{equation}
Thus, we have diagonalized the lapse sector.
\subsubsection*{Shift sector}
For the shift $N_i$ we do the same. We collect all the terms containing $N_i$ coming from Eqs. \eqref{quadraticactionmomentaonlydiagonalised} and \eqref{quadraticactionlapsediagonalised} and the $-N_i\mathcal{H}^{i(1)}$ term in Eq. \eqref{quadraticnonminimalaction}. We can now write
\begin{equation}
S_{N_i}^{(2)}=\int d^Dx \left\{\frac{a^{D-5}F}{\bar{N}}\left(\left[\partial_{\left(i\right.}N_{\left.j\right)}\right]^2-\alpha (\partial_iN^{i})^2+ J_{ij}\partial^{\left(i\right.}N^{\left.j\right)} \right)\right\},
\label{quadraticactionshiftonly}
\end{equation}
where $|\partial_{\left(i\right.}N_{\left.j\right)}|^2=\partial_{\left(i\right.}N_{\left.j\right)}\partial^{\left(i\right.}N^{\left.j\right)}$ and $(\partial_iN^{i})^2=\partial_iN^{i}\partial_jN^{j}$. Furthermore,
\begin{align}
\alpha=&\frac{1}{D-1}\left(\frac{\frac{1}{F}\frac{1}{a^{D-3}}\frac{\mathcal{P}^2}{4(D-1)(D-2)}-(D-1)\frac{\hat{\mathcal{P}}_{\phi}^2}{2\bar{\Omega} a^{D-1}}}{-W}\right)\label{definitionalpha}\\
\nonumber &\xrightarrow{\text{on-shell}} \frac{1}{D-1}\left(1-\frac{(D-2)\hat{\mathcal{P}}_{\phi}^2}{2\bar{\Omega} a^{2(D-1)}V}\right)= \frac{1}{D-1}\left(1-\frac{(D-2)\bar{\Omega}\dot{\phi}^2}{2V}\right)
\end{align}
and
\begin{align}
\nonumber J_{ij}=&\bar{N}a^2\Biggl\{-\dot{h}_{ij}+J_{h_{ij}}h_{ij}\\
&+\delta_{ij}\Biggl(\alpha \dot{h}-\frac{\mathcal{P}}{2(D-1)a W}\left[\partial_k\partial_{l}h^{kl}-\partial_k\partial^{k} h\right] +\frac{\mathcal{P}}{(D-1)a W}\frac{F'}{F}a^{D-3}\partial_k\partial^k\varphi
+J_{\dot{\varphi}}\dot{\varphi}+J_{\varphi}\varphi+J_h h\Biggr)\Biggr\},
\end{align}
where
\begin{align}
J_{h_{ij}}=&-\frac{\mathcal{P}}{(D-1)(D-2)a^{D-2}F}-2\left[H+\frac{1}{D-2}\frac{F'}{F}\dot{\phi}\right]\xrightarrow{\text{on-shell}} 0\\
J_{\dot{\varphi}}=&2\frac{F'}{F}+\frac{1}{2(D-1)F a^{D-2}}\frac{\mathcal{P}}{W}\left(\hat{\mathcal{P}}_{\phi}+\frac{a\mathcal{P}F'}{(D-2)F}\right)=\xrightarrow{\text{on-shell}} \frac{\bar{\Omega}}{V}(D-2)H\dot{\phi}\\
J_{\varphi}=&2\left(\frac{F'}{F}\right)'\dot{\phi}+\frac{\hat{\mathcal{P}}_{\phi}}{a^{D-1}F}+\frac{\mathcal{P}}{(D-1)(D-2)a^{D-2}}\frac{F'}{F^2} -\frac{\mathcal{P}}{2(D-1)a^{D-2}F}\frac{C_{\varphi}}{W}\\
\nonumber&\xrightarrow{\text{on-shell}}\frac{1}{F}\left[(1+2F'')\dot{\phi}-2F'H-\frac{(D-2)FH+F'\dot{\phi}}{a^{D-1}V}C^{\text{on-shell}}_{\varphi}\right]\\
\nonumber & =a^{D-1}2(D-1)(D-2)\frac{\bar{\Omega} H^2}{W} \left(\frac{\dot{\phi}}{2(D-1)H}\right)^{\cdot}\\
J_h=&\frac{\mathcal{P}}{(D-1)(D-2)a^{D-2}F}+2\left[H+\frac{1}{D-2}\frac{F'}{F}\dot{\phi}\right]+\frac{\mathcal{P}}{4(D-1)a^{D-2}}\frac{C_h}{W}\xrightarrow{\text{on-shell}} 0
,
\end{align}
where $C_{\varphi}$ and $C_h$ were defined in Eqs. \eqref{nshiftvarphi} and \eqref{nshifth}. In order to diagonalize the shift sector \eqref{quadraticactionshiftonly}, we separate the longitudinal and transversal degrees of the shift vector,
\begin{equation}
N_i=\partial_i S+N_{i}^T,~~~~~~~~~~~~~~~~\text{with}~~~~~~~~~~\partial^{i}N_i^T=0.
\label{shiftdecomposition2}
\end{equation}
Then, up to boundary terms, the shift action \eqref{quadraticactionshiftonly} becomes
\begin{equation}
S_{N_i}^{(2)}=\int d^Dx \left\{\frac{a^{D-5}F}{\bar{N}}\left((1-\alpha)\left[\nabla^2 S\right]^2+[\partial_{\left(i\right.}N^T_{\left.j\right)}]^2+ J_{ij}\partial^{i}\partial^{j}S -\bar{N}a^2\left(\dot{h_{ij}}-J_{h_{ij}}h_{ij}\right)\partial^{\left(i\right.}N^{T\left.j\right)} \right)\right\}.
\label{quadraticactionshiftonlydecomposed}
\end{equation}
Note that we have been able to separate the longitudinal and transversal degrees of the shift vector. Now we substitute the scalar-vector-tensor decomposition of $h_{ij}$, see \eqref{scalarvectortensordecomposition}. With this decomposition, we can show that
\begin{align*}
J_{ij}\partial^{i}\partial^{j}S=&\frac{1}{D-1}\left[J-(D-2)\bar{N}a^2\nabla^2 (\dot{\tilde{h}}-J_{h_{ij}}\tilde{h})\right]\nabla^2 S\\
-\bar{N}a^2\dot{h_{ij}}\partial^{\left(i\right.}N^{T\left.j\right)}=&-\bar{N}a^2\left(\partial_{\left(i\right.}\dot{h}^T_{\left.j\right)} -J_{h_{ij}}\partial_{\left(i\right.}h^T_{\left.j\right)}\right)\partial^{\left(i\right.}N^{T\left.j\right)},
\end{align*}
where $J=\delta^{ij}J_{ij}$ and we did not write down any total derivative terms. We can now diagonalize the action \eqref{quadraticactionshiftonlydecomposed} by introducing
\begin{align}
\nabla^2\tilde{S}=&\nabla^2S+\frac{1}{2(D-1)}\frac{1}{1-\alpha}\left[J-(D-2)\bar{N}a^2\nabla^2 (\dot{\tilde{h}}-J_{h_{ij}}\tilde{h})\right]\label{tildeS}\\
\partial_{\left(i\right.}\tilde{N}^T_{\left.j\right)}=&\partial_{\left(i\right.}N^T_{\left.j\right)}-\frac{a^2\bar{N}}{2}\left(\partial_{\left(i\right.}\dot{h}^T_{\left.j\right)} -J_{h_{ij}}\partial_{\left(i\right.}h^T_{\left.j\right)}\right),\label{tildeNiT}
\end{align}
and the action becomes
\begin{align}
\nonumber S_{N_i}^{(2)}=&\int d^Dx\bar{N} \Biggl\{\frac{a^{D-5}F}{\bar{N}^2}\left((1-\alpha)\left[\nabla^2 \tilde{S}\right]^2+[\partial_{\left(i\right.}\tilde{N}^T_{\left.j\right)}]^2\right) -\frac{a^{D-1}F}{4}[\partial_{\left(i\right.}\dot{h}^T_{\left.j\right)} -J_{h_{ij}}\partial_{\left(i\right.}h^T_{\left.j\right)}]^2\\
&-\frac{a^{D-5}F}{\bar{N}^2}\frac{1}{4(D-1)^2}\frac{1}{1-\alpha}\left[J-(D-2)\bar{N}a^2\nabla^2 (\dot{\tilde{h}}-J_{h_{ij}}\tilde{h})\right]^2\Biggr\}.
\label{quadraticactionshiftdiagonalized}
\end{align}

\subsubsection*{Gauge invariant scalar action}
So far we have diagonalized the perturbed momenta $\pi^{ij}$ and $\hat{\pi}_{\phi}$ and the perturbed lapse $n$ and shift $N_i$ functions. What remains from the action are the field and metric perturbations $\varphi$ and $h_{ij}$. For the interested reader who wants to check the derivation we present an intermediate result where we collect all the terms containing these fields. Since there are many terms containing these fields we will do this step by step. First of all we collect the leftover terms in the action \eqref{quadraticnonminimalaction}
\begin{align}
\nonumber S^{(2)}_{h,\varphi}=\int d^Dx \frac{\bar{N}a^{D-1}F}{4}&\biggl\{\frac{4\mathcal{P}}{(D-1)(D-2)a^{D-2}F}\biggl[\frac{F'}{F}h\dot{\varphi}+\frac12 (D-1)\left(\frac{F'}{F}\right)''\varphi^2\dot{\phi} +\left(\frac{F'}{F}\right)'\varphi\left((D-1)\dot{\varphi}+h\dot{\phi}\right)\biggr]\\
\nonumber &+\frac{4}{a^2}\left[-\frac14 h\nabla^2h+\frac12 h\partial^{i}\partial^{j}h_{ij}-\frac12 h_{ij}\partial^{i}\partial^{l}h_{jl}+\frac14 h^{ij}\nabla^2h_{ij}+\frac{F'}{F}\varphi(\partial^{i}\partial^{j}h_{ij}-\nabla^2h)\right]\\
\nonumber & -\left[-\frac{\mathcal{P}^2}{4(D-1)(D-2)a^{D-3}F}+\frac{\tilde{\mathcal{P}}_{\phi}^2}{2\bar{\Omega} a^{D-1}}+a^{D-1}V\right]\frac{\frac12 h^2}{a^{D-1}F}\\
\nonumber & -\left[\frac{\mathcal{P}^2}{4(D-1)(D-2)a^{D-3}F}\frac{3D-7}{D-1}+\frac{\tilde{\mathcal{P}}_{\phi}^2}{2\bar{\Omega} a^{D-1}}-a^{D-1}V\right]\frac{h_i^jh_j^{i}}{a^{D-1}F} \\
\nonumber & -\left[-\frac{\mathcal{P}^2}{4(D-1)(D-2)a^{D-3}}\left(\frac{1}{F}\right)''+\frac{\tilde{\mathcal{P}}_{\phi}^2}{2 a^{D-1}}\left(\frac{1}{\bar{\Omega}}\right)''+a^{D-1}V_{,\phi\phi}\right]\frac{2 \varphi^2}{a^{D-1}F} \\
\nonumber &-\left[\frac{\mathcal{P}^2}{4(D-1)(D-2)a^{D-3}}\left(\frac{1}{F}\right)'\frac{D-5}{D-1}-\frac{\tilde{\mathcal{P}}_{\phi}^2}{2a^{D-1}}\left(\frac{1}{\bar{\Omega}}\right)'+a^{D-1} V_{,\phi}\right]\frac{2 h\varphi}{a^{D-1}F} \\
&-\frac{1}{a^2 F} 2 \partial^{i}\varphi\partial_i\varphi\biggr\}.
\label{quadraticnonminimalactionhvarphi}
\end{align}
Remember that we diagonalized the momenta by shifting them. By doing this we managed to write all the momentum terms from the action in terms of quadratic shifted momenta, see Eq. \eqref{quadraticactionmomentaonlydiagonalised}. As a consequence, we obtained additional terms that do not contain the momenta, but do contain $h$ and $\varphi$ and combinations. We find
\begin{align}
\nonumber \frac{\bar{N}\tilde{\mathcal{P}}_{\phi}^2}{2a^{D-1}\bar{\Omega}}\frac14 \hat{I}_{\varphi}^2(h,\varphi)&=\frac{\bar{N}a^{D-1}F}{4}\Biggl\{\frac{\tilde{\mathcal{P}}_{\phi}^2}{a^{D-1}\bar{\Omega}}\frac{\frac12 h^2}{a^{D-1}F}+\frac{\bar{\Omega}}{F}2\dot{\varphi}^2 +\frac{2\tilde{\mathcal{P}}_{\phi}^2\left(\bar{\Omega}'\right)^2}{F \bar{\Omega}^3(a^{D-1})^2}\varphi^2\\
& +\frac{\tilde{\mathcal{P}}_{\phi}}{a^{D-1}F}2h\dot{\varphi}+\frac{2\tilde{\mathcal{P}}_{\phi}^2\bar{\Omega}'}{F \bar{\Omega}^2(a^{D-1})^2}h\varphi+\frac{4\tilde{\mathcal{P}}_{\phi}\bar{\Omega}'}{a^{D-1}\bar{\Omega} F}\dot{\varphi}\varphi\Biggr\}.
\label{Ivarphi2contribution}
\end{align}
Now we consider the other part of Eq. \eqref{quadraticactionmomentaonlydiagonalised} with the term $I_{ij}I^{ij}-I^2$. For this term we will put the canonical momentum $\mathcal{P}$ on-shell in the term multiplying $h_{ij}$ in Eq. \eqref{shiftinpiij}. This will dramatically simplify our calculations. We find
\begin{align}
\nonumber &\frac{\mathcal{P}^2}{4(D-1)^2a^{D-3}F}\frac14 \left(I_{ij}I^{ij}-I^2\right)=\\
\nonumber \frac{\bar{N}a^{D-1}F}{4}&\Biggl\{ \dot{h_{ij}}\dot{h^{ij}}-\dot{h}^2-\frac{2\mathcal{P}}{(D-1)Fa^{D-2}}\left[h^{ij}\dot{h_{ij}}-\frac12 h \dot{h}\right]\\
\nonumber &+\frac{\mathcal{P}^2}{4(D-1)(D-2)a^{D-3}F}\frac{\left[4\frac{D-2}{D-1}h_i^jh_j^{i}-h^2\right]}{a^{D-1}F}-4\frac{D-1}{D-2}\frac{F^{\prime 2}}{F^2}\dot{\varphi}^2\\
\nonumber &-\left[4\frac{D-1}{D-2}\left(\frac{F'}{F}\right)^{\prime 2}\dot{\phi}^2+\frac{\mathcal{P}^2}{(D-1)(D-2)a^{2(D-2)}}\left(\frac{1}{F}\right)^{\prime 2} +\frac{4 \mathcal{P}\dot{\phi}}{(D-2)a^{D-2}}\left(\frac{F'}{F}\right)'\left(\frac{1}{F}\right)'\right]\varphi^2\\
\nonumber &-\left[\frac{2\mathcal{P}}{a^{D-2}(D-2)}\frac{F'}{F^2}\left(\frac{4(D-1)Fa^{D-2}}{\mathcal{P}}\left(\frac{F'}{F}\right)'\dot{\phi}+2F\left(\frac{1}{F}\right)'\right)\right]\varphi\dot{\varphi}\\
\nonumber &-4\frac{F'}{F} \dot{h}\dot{\varphi}-\left[4\left(\frac{F'}{F}\right)'\dot{\phi}+\frac{2\mathcal{P}}{(D-1)a^{D-2}}\left(\frac{1}{F}\right)'\right]\dot{h}\varphi +\frac{2(D-3)\mathcal{P}}{(D-1)(D-2)a^{D-2}F}\frac{F'}{F}h\dot{\varphi}\\
& +\left[\frac{2(D-3)\mathcal{P}}{(D-1)(D-2)a^{D-2}F}\left(\frac{F'}{F}\right)'\dot{\phi}+\frac{(D-3)\mathcal{P}^2}{(D-1)^2(D-2)a^{2(D-2)}F}\left(\frac{1}{F}\right)'\right]h\varphi\Biggr\}.
\label{Iij2contribution}
\end{align}
The completion of the square for $n$ also gives extra terms,
\begin{align}
\nonumber &-\frac{\bar{N}}{4 W}I_n^2=\\
\nonumber \frac{\bar{N}a^{D-1}F}{4}\frac{1}{W}&\Biggl\{ -\frac{\mathcal{P}^2}{4(D-1)^2a^{D-3}F}\dot{h}^2-\frac{\mathcal{P}}{2(D-1)a^{D-2}F}C_h h \dot{h}-\frac{1}{4F a^{D-1}}C_h^2h^2\\
\nonumber&-\frac{1}{F a^{D-1}}\left(\tilde{\mathcal{P}}_{\phi}+\frac{a \mathcal{P} F'}{(D-2)F}\right)^2\dot{\varphi}^2+\frac{2C_{\varphi}}{F a^{D-1}}\left(\tilde{\mathcal{P}}_{\phi}+\frac{a \mathcal{P} F'}{(D-2)F}\right)\varphi\dot{\varphi}-\frac{1}{F a^{D-1}}C_{\varphi}^2\varphi^2\\
\nonumber & -\frac{\mathcal{P}}{(D-1)a^{D-2}F}\left(\tilde{\mathcal{P}}_{\phi}+\frac{a \mathcal{P} F'}{(D-2)F}\right)\dot{h}\dot{\varphi}+\frac{\mathcal{P}}{(D-1)a^{D-2}F}C_{\varphi}\dot{h}\varphi\\ \nonumber & -\frac{1}{F a^{D-1}}\left(\tilde{\mathcal{P}}_{\phi}+\frac{a \mathcal{P} F'}{(D-2)F}\right)C_h h\dot{\varphi}+\frac{1}{F a^{D-1}}C_{h}C_{\varphi}h\varphi\\
\nonumber & -\frac{2}{a^2}\left[\partial_i\partial_{j}h^{ij}-\partial_i\partial^{i} h\right]\left(-\frac{a\mathcal{P}}{2(D-1)}\dot{h}-C_h\frac{h}{2}-\left(\tilde{\mathcal{P}}_{\phi}+\frac{a\mathcal{P}F'}{(D-2)F}\right)\dot{\varphi}+C_{\varphi}\varphi\right)\\
\nonumber & +\frac{4 F'}{F a^{2}}\nabla^2\varphi \left(-\frac{a\mathcal{P}}{2(D-1)}\dot{h}-C_h\frac{h}{2}-\left(\tilde{\mathcal{P}}_{\phi}+\frac{a\mathcal{P}F'}{(D-2)F}\right)\dot{\varphi}+C_{\varphi}\varphi\right)\\
& -\frac{1}{F a^{D-1}}\left(a^{D-3}F\left[\partial_i\partial_{j}h^{ij}-\partial_i\partial^{i} h\right]-2a^{D-3}F'\nabla^2\varphi\right)^2 \Biggr\}.
\label{In2contribution}
\end{align}
As for the final contributions coming from completing the square in the shift sector, we use the following useful relations:
\begin{align}
\frac{1}{1-\alpha}&=\frac{-W}{\frac{\mathcal{P}^2}{4(D-1)^2a^{D-3}F}}\\
(D-1)\alpha-1&=\frac{D-2}{W}\frac{\tilde{\mathcal{P}}_{\phi}^2}{2\bar{\Omega} a^{D-1}},
\end{align}
which gives
\begin{align}
\nonumber & -\bar{N}\frac{a^{D-5}F}{\bar{N}^2}\frac{1}{4(D-1)^2}\frac{1}{1-\alpha}\left[J-(D-2)\bar{N}a^2\nabla^2 (\dot{\tilde{h}}-J_{h_{ij}}\tilde{h})\right]^2=\\
\nonumber  \frac{\bar{N}a^{D-1}F}{4}\frac{W}{\frac{\mathcal{P}^2}{4a^{D-3}F}}&\Biggl[ \frac{D-2}{W}\frac{\tilde{\mathcal{P}}_{\phi}^2}{2\bar{\Omega} a^{D-1}}\dot{h}-(D-2)\nabla^2 \dot{\tilde{h}}+\left[J_{h_{ij}}+(D-1)J_h\right]h +(D-2)J_{h_{ij}}\nabla^2\tilde{h}\\
& -\frac{\mathcal{P}}{2 a W}\left[\partial_i\partial_{j}h^{ij}-\partial_i\partial^{i} h\right]+\frac{\mathcal{P}}{a^{D-2} W}\frac{F'}{F}a^{D-3}\nabla^2\varphi +(D-1)J_{\dot{\varphi}}\dot{\varphi} +(D-1)J_{\varphi}\varphi\Biggr]^2.
\label{INicontribution}
\end{align}
Now that we have obtained all the terms containing $h$ and $\varphi$, we want to express the scalar part of Eqs. \eqref{quadraticnonminimalactionhvarphi},  \eqref{Ivarphi2contribution}, \eqref{Iij2contribution}, \eqref{In2contribution} and \eqref{INicontribution} in terms of the gauge invariant variable
\begin{equation}
\tilde{\varphi}=\varphi-\frac{1}{2(D-1)}\frac{\dot{\phi}}{H}(h-\nabla^2\tilde{h}).\label{tildevarphi}
\end{equation}
This is a tedious procedure which we will not explicitly write down in this Appendix. However, we can provide the interested reader with calculations and intermediate steps. An important step in the derivation is a partial integration of all the terms containing only one time derivative. The final diagonalized free action we obtain is
\begin{align}
\nonumber S^{(2)}=&\int d^{D-1}x \bar{N}dt a^{D-1}\Biggl\{\frac{z^2}{z_0^2}\left[\frac12\dot{\tilde{\varphi}}^2-\frac12 \left(\frac{\partial_i\tilde{\varphi}}{a}\right)^2+\frac12 \frac{z_0}{a^{D-1} z^2}\left[a^{D-1}\frac{z^2}{z_0^2}\dot{z_0}\right]^{\cdot}\tilde{\varphi}^2\right]+\frac{F}{4}\biggl[(\dot{h}_{ij}^{TT})^2-\Bigl(\frac{\partial h_{ij}^{TT}}{a}\Bigr)^2\biggr]\\
&-\frac{\hat{\mathcal{P}}_{\phi}^2}{2a^{2(D-1)}\bar{\Omega}}\hat{\tilde{\pi}}_{\varphi}^2-\frac{\mathcal{P}^2}{4(D-1)^2a^{2(D-2)}F}\tilde{\pi}^{ij} \frac{A_{ijkl}}{2}\tilde{\pi}^{kl}+\frac{W}{a^{D-1}\bar{N}^2}\tilde{n}^2+\frac{F}{a^4 \bar{N}^2}\left((1-\alpha)[\nabla^2 \tilde{S}]^2+[\partial_{\left(i\right.}\tilde{N}^T_{\left.j\right)}]^2\right)\Biggr\},
\label{freeActionappendix}
\end{align}
where
\begin{align}
z_0^2&=\frac{1}{4(D-1)^2}\frac{\dot{\phi}^2}{H^2}\\
z^2&=\frac{1}{4(D-1)^2}\frac{\bar{\Omega} \dot{\phi}^2}{(H+\frac{1}{D-2}\frac{F'}{F}\dot{\phi})^2}.
\end{align}
Interestingly enough we only have to use the on-shell Friedmann and field equations \eqref{nonminimalH2}-\eqref{fieldequation} for the single time derivative terms after we have partially integrated these terms. Therefore the free action \eqref{freeActionappendix} is obtained up to some terms multiplying the on-shell relations. Since the-off-shell corrections are only higher order in perturbations, we can neglect these terms.\\
The action \eqref{freeActionappendix} is time parametrization invariant since the function $\bar{N}$ can be picked arbitrarily. For example, if $\bar{N}=1$ corresponds to real time, whereas we switch to conformal time $d\eta=a dt$ if we choose $\bar{N}=a$. In conformal time we can define another convenient variable, the (generalized) Mukhanov-Sasaki variable
\begin{equation}
v=a^{\frac{D-2}{2}}\frac{z}{z_0}\tilde{\varphi},
\end{equation}
which changes the scalar part of the action to
\begin{align}
S_v^{(2)}=\int d^{D-1}x d\eta \frac12 \left[v^{\prime 2}-(\partial_i v)^2+\frac{\bigl(a^{\frac{D-2}{2}} z\bigr)''}{a^{\frac{D-2}{2}}z}v^2\right].
\end{align}
Another convenient variable is the comoving curvature perturbation, which has the special property that it is invariant under a conformal transformation. In the main text we show how the action transforms in terms of this variable.

\section{\label{sec:appendixgaugeinvariance}Gauge invariance of the dynamical and constraint fields in the free action}
In this Appendix we want to verify that the action \eqref{freeAction} is gauge invariant (up to linear order in coordinate transformations). The metric transforms under an infinitesimal coordinate transformation $x^{\mu}\rightarrow x^{\mu}+\xi^{\mu}(x)$ as $g_{\mu\nu}\rightarrow g_{\mu\nu}-\nabla_{\mu}\xi_{\nu}-\nabla_{\nu}\xi_{\mu}$, $\xi_{\mu}=g_{\mu\nu}\xi^{\nu}$. The metric components to linear order in perturbations are
\begin{align}
g_{00}^{(1)}=-2\bar{N}n; \qquad g_{0i}^{(1)}=N_i;  \qquad  g_{ij}^{(1)}=a^2 h_{ij},
\end{align}
whereas the nonvanishing background Christoffel connections are,
\begin{align}
\Gamma^{(0)0}_{00}=\dot{\bar{N}}; \qquad \Gamma^{(0)i}_{0j}=\bar{N} H \delta^{i}_{j};  \qquad  \Gamma^{(0)0}_{ij}=\frac{a^2 H}{\bar{N}}\delta_{ij}.
\end{align}
If we now decompose $h_{ij}$ as in Eqs. \eqref{scalarvectortensordecomposition} and \eqref{shiftdecomposition}, and similarly $\xi_i=\xi_i=\xi_i^T+\partial_i\xi$, $\partial_i\xi_i^T=0$, we find that the perturbed metric fields transform as
\begin{align}
\nonumber &n~ \rightarrow ~n+\dot{\xi}_0-\frac{\dot{\bar{N}}}{\bar{N}}\xi_0; \qquad N_i^T~\rightarrow~ N_i^T -\bar{N}\dot{\xi}_i^T+2 \bar{N} H\xi_i^T; \qquad S~\rightarrow ~S-\bar{N}\dot{\xi}-\xi_0+2\bar{N}H \xi\\
&h~\rightarrow ~h-2\frac{\nabla^2\xi}{a^2}+\frac{2(D-1)H}{\bar{N}}\xi_0; \qquad \tilde{h}~\rightarrow ~\tilde{h} -\frac{2}{a^2}\xi;\qquad h_i^T~\rightarrow ~h_i^T-\frac{2}{a^2}\xi_i^T; \qquad h_{ij}^{TT}~\rightarrow ~h_{ij}^{TT}.\label{transformationsguv}
\end{align}
Moreover, from $\Phi(x^{\mu})\rightarrow \Phi(x^{\mu}+\xi^{\mu})$ we find that
\begin{equation}
\varphi~\rightarrow~\varphi+\frac{\dot{\phi}}{\bar{N}}\xi_0.\label{transformationvarphi}
\end{equation}
By substituting the transformations \eqref{transformationsguv} and \eqref{transformationvarphi} into the shifted fields \eqref{tilden}, \eqref{tildeS}, \eqref{tildeNiT} and \eqref{tildevarphi} we find that $\tilde{n},~\tilde{S},~\tilde{N}_i^T,~\tilde{\varphi}$ and $h_{ij}^{TT}$ are invariant under linear coordinate transformations.\\
The free action \eqref{freeAction} is only completely gauge invariant if the canonical momenta $\hat{\tilde{\pi}}_{\varphi}$ and $\tilde{\pi}^{ij}$ are also gauge invariant. We shall now argue why this is so. For this we go back to the canonical form of the action for a nonminimally coupled scalar field, \textit{i.e.} Eq. \eqref{canonicalactionnonminimalshifted2}. Remember that we derived this action from the original action \eqref{nonminimalaction}. In other words, if we vary the action \eqref{canonicalactionnonminimalshifted2} with respect to the canonical momenta and substitute these back into action, we obtain the diffeomorphism invariant action \eqref{nonminimalaction}. Since this action is invariant under linear coordinate transformations, this means the canonical momenta should also transform in such a way that the free action \eqref{freeAction} is gauge invariant.\\
To see how this works, let us vary the free action \eqref{freeAction} with respect to $\hat{\tilde{\pi}}_{\varphi}$ and $\tilde{\pi}^{ij}$. We find that $\hat{\tilde{\pi}}_{\varphi}=\hat{\pi}_{\varphi}+\hat{I}_{\varphi}/2=0$ and $\tilde{\pi}^{ij}=\pi^{ij}+(I_{ij}-\delta_{ij}I)/2=0$, which are the Hamilton equations of the linearized theory and must be gauge invariant. Therefore $\hat{\tilde{\pi}}_{\varphi}$ and $\tilde{\pi}^{ij}$ are gauge invariant, and the free action \eqref{freeAction} as a whole is gauge invariant. From the transformations \eqref{transformationsguv} and \eqref{transformationvarphi} and Eqs. \eqref{shigtinpiphi} and \eqref{shiftinpiij} we find that
\begin{align}
\hat{I}_{\varphi}~~\rightarrow~~ & \hat{I}_{\varphi}+2\frac{\nabla^2\xi}{a^2}-\frac{2}{\bar{N}\dot{\phi}}(\ddot{\phi}+(D-1)H\dot{\phi})\xi_0-\frac{2}{\bar{N}}\frac{\dot{\bar{\Omega}}}{\bar{\Omega}}\xi_0\\
\nonumber I_{ij}~~\rightarrow~~ &I_{ij}-\frac{4}{a^2}\partial_{\left( i\right.}\xi_{\left. j\right)}+\frac{2}{D-2}\frac{\nabla^2 \xi}{a^2}\delta_{ij}+\frac{2}{D-2}\frac{1}{H+\frac{1}{D-2}\frac{F'}{F}\dot{\phi}}\frac{\partial_i\partial_j\xi_0}{\bar{N}a^2}\\
&+\frac{2}{D-2}\left[\frac{\left(H+\frac{1}{D-2}\frac{F'}{F}\dot{\phi}\right)^{\cdot}}{H+\frac{1}{D-2}\frac{F'}{F}\dot{\phi}}+(D-3)H+\frac{F'}{F}\dot{\phi}\right]\frac{\delta_{ij}\xi_0}{\bar{N}},
\end{align}
such that the momenta transform as
\begin{align}
\hat{\pi}_{\varphi}~\rightarrow~& \hat{\pi}_{\varphi}-\frac{\nabla^2\xi}{a^2}+\frac{1}{\bar{N}\dot{\phi}}(\ddot{\phi}+(D-1)H\dot{\phi})\xi_0+\frac{1}{\bar{N}}\frac{\dot{\bar{\Omega}}}{\bar{\Omega}}\xi_0\\
\nonumber \pi^{ij}~\rightarrow~ &\pi^{ij}+\frac{2}{a^2}\partial_{\left( i\right.}\xi_{\left. j\right)}-\frac{\nabla^2 \xi}{a^2}\delta_{ij}-\frac{2}{D-2}\frac{1}{H+\frac{1}{D-2}\frac{F'}{F}\dot{\phi}}\frac{(\partial_i\partial_j-\delta_{ij}\nabla^2)\xi_0}{\bar{N}a^2}\\
&+\left[\frac{\left(H+\frac{1}{D-2}\frac{F'}{F}\dot{\phi}\right)^{\cdot}}{H+\frac{1}{D-2}\frac{F'}{F}\dot{\phi}}+(D-3)H+\frac{F'}{F}\dot{\phi}\right]\frac{\delta_{ij}\xi_0}{\bar{N}}.
\end{align}

\section{\label{sec:appendixconformalinvariance} Conformal invariance of the dynamical and constraint fields in the free action}
In the main text we showed that the dynamical and constraint fields in the free action \eqref{freeAction} should transform as Eqs. \eqref{conformaltransfn}-\eqref{conformaltransfhij} if the free Jordan frame action is physically equivalent to the free action in the Einstein frame. In this Appendix we prove that this is indeed the case. For the dynamical scalar field in \eqref{freeAction} we already proved that the Jordan and Einstein frame actions are physically equivalent. Our approach will be similar to Appendix \ref{sec:appendixgaugeinvariance} where we showed gauge invariance of the free action.\\
We first consider the unperturbed action \eqref{canonicalactionnonminimalshifted2}. Under a conformal transformation the metric transforms as $g_{\mu\nu,E}=\omega(\Phi)^2 g_{\mu\nu}$ with the identification $\omega^{D-2}=F(\Phi)$. The action \eqref{canonicalactionnonminimalshifted2} can be transformed to the Einstein frame action through the following relations:
\begin{align}
\nonumber &N_E=\omega N \qquad N_{i,E}=\omega^2 N_i \qquad g_{ij,E}=\omega^2g_{ij}\\
&p^{ij}_E=\frac{1}{\omega^2} p^{ij} \qquad p_{\Phi,E}=\sqrt{\frac{\omega^{D-2}}{\Omega}}\hat{p}_{\Phi}   \qquad \partial_{\mu}\Phi_E=\sqrt{\frac{\Omega}{\omega^{D-2}}}\partial_{\mu}\Phi \qquad V_E=\frac{1}{\omega^D}V.\label{relationsEinsteinJordanframe}
\end{align}
In this equations the quantity $\Omega$ (see Eq. \eqref{definitionOmega}) is understood to be unperturbed, \textit{i.e.} $\Omega=\Omega(\Phi)$. We now decompose the metric as $g_{\mu\nu}=\bar{g}_{\mu\nu}+\delta g_{\mu\nu}$, where the perturbed metric contains the lapse, shift and spatial metric perturbations. Similarly we substitute $\Phi=\phi+\varphi$, thereby decomposing the conformal factor, as well as the factor $\Omega$, in a background and perturbed part, $\omega=\bar{\omega}+\delta\omega$. The background fields and momenta then transform as in Eqs. \eqref{relationlapse}, \eqref{relationscalefactor}, \eqref{relationEinsteinframemmomentumP}-\eqref{relationpotential}, \eqref{relationHubble} and \eqref{relationdotphi}. It is easy to show that the perturbed metric fields transform as
\begin{align}
\nonumber &n_E=\bar{\omega} \left(n+\bar{N}\frac{\delta\omega}{\bar{\omega}}\right); \qquad N_{i.E}^T=\bar{\omega}^2N_i^T; \qquad S_E=\bar{\omega}^2 S\\
&h_E=h+2(D-1)\frac{\delta\omega}{\bar{\omega}};\qquad \tilde{h}_E=\tilde{h}; \qquad h_{i,E}^T=h_i^T; \qquad h_{ij,E}^{TT}=h_{ij}^{TT},
\end{align}
where $\frac{\delta\omega}{\bar{\omega}}= \frac{1}{D-2}\frac{F'}{F}\varphi$. From here we can already see that the graviton is invariant under a conformal transformation. Furthermore, from Eqs. \eqref{relationvarphi} and \eqref{relationdotphi} we find that
\begin{equation}
\varphi_E=\sqrt{\frac{\bar{\Omega}}{\bar{\omega}^{D-2}}}\varphi.
\label{relationvarphi2}
\end{equation}
We now substitute these relations into the Einstein frame fields $\tilde{n}_E$, $\tilde{S}_E$ and $\tilde{N}_{i,E}^T$, which we obtain from Eqs. \eqref{tilden}, \eqref{tildeS} and \eqref{tildeNiT} by setting $F=1$. We find indeed that these fields transform under the conformal transformation as, $\tilde{n}_E=\bar{\omega}\tilde{n}$, $\tilde{S}_E=\bar{\omega}^2\tilde{S}$ and $\tilde{N}^T_{i,E}=\bar{\omega}^2\tilde{N}^T_{i}$. Therefore we conclude that the corresponding terms in the free Einstein frame action \eqref{freeActionEinstein} are physically equivalent to the terms in the Jordan frame action \eqref{freeAction}.\\
We now want to show that the shifted momentum perurbations $\tilde{\pi}_{\varphi,E}$ and $\tilde{\pi}_E^{ij}$ are invariant under a conformal transformation. Consider the unperturbed Einstein frame momenta $p_{\Phi,E}$ and $p^{ij}_E$. If we use the relation between the Einstein and Jordan frame momenta from Eq. \eqref{relationsEinsteinJordanframe} and expand up to linear order in perturbations, we find that the momentum perturbations are related as
\begin{align}
\pi_{\varphi,E}&=\hat{\pi}_{\varphi}+\frac12 \left(\frac{F'}{F}-\frac{\bar{\Omega}'}{\bar{\Omega}}\right)\varphi\\
\pi^{ij}_E&=\pi^{ij}-\frac{2}{D-2}\frac{F'}{F}\varphi \delta^{ij}.
\end{align}
On the other hand we can also perform the conformal transformation of the shifts in the momenta, $I_{\varphi,E}$ and $I_{ij,E}$ from Eqs. \eqref{shigtinpiphi} and \eqref{shiftinpiij}. We find that under the conformal transformation
\begin{align}
I_{\varphi,E}&= \hat{I}_{\varphi}+\left(\frac{\bar{\Omega}'}{\bar{\Omega}}-\frac{F'}{F}\right)\varphi\\
I_{ij,E}&=I_{ij}-\frac{4}{(D-2)^2}\frac{F'}{F}\varphi\delta_{ij}.
\end{align}
Therefore the shifted momentum perturbations $\tilde{\pi}_{\varphi,E}=\pi_{\varphi,E}+\frac12 I_{\varphi,E}$ and $\tilde{\pi}_E^{ij}=\pi_E^{ij}+\frac12 \left(I^{E}_{ij}-\delta_{ij} I^{E}\right)$ transform as
\begin{align}
&\tilde{\pi}_{\varphi,E}\rightarrow \hat{\tilde{\pi}}_{\varphi}\\
&\tilde{\pi}_E^{ij}\rightarrow \tilde{\pi}^{ij}.
\end{align}
Therefore the corresponding terms in the free Einstein action \eqref{freeActionEinstein} are physically equivalent to those in the Jordan frame. In conclusion, we prove the physical equivalence of the free Jordan and Einstein frame actions for all dynamical and constraint fields.

\bibliography{nonminimalADM}{}

\begin{thebibliography}{10}%
\makeatletter
\providecommand \@ifxundefined [1]{%
 \ifx #1\undefined \expandafter \@firstoftwo
 \else \expandafter \@secondoftwo
\fi
}%
\providecommand \@ifnum [1]{%
 \ifnum #1\expandafter \@firstoftwo
 \else \expandafter \@secondoftwo
\fi
}%
\providecommand \enquote [1]{``#1''}%
\providecommand \bibnamefont  [1]{#1}%
\providecommand \bibfnamefont [1]{#1}%
\providecommand \citenamefont [1]{#1}%
\providecommand\href[0]{\@sanitize\@href}%
\providecommand\@href[1]{\endgroup\@@startlink{#1}\endgroup\@@href}%
\providecommand\@@href[1]{#1\@@endlink}%
\providecommand \@sanitize [0]{\begingroup\catcode`\&12\catcode`\#12\relax}%
\@ifxundefined \pdfoutput {\@firstoftwo}{%
 \@ifnum{\z@=\pdfoutput}{\@firstoftwo}{\@secondoftwo}%
}{%
 \providecommand\@@startlink[1]{\leavevmode\special{html:<a href="#1">}}%
 \providecommand\@@endlink[0]{\special{html:</a>}}%
}{%
 \providecommand\@@startlink[1]{%
  \leavevmode
  \pdfstartlink
   attr{/Border[0 0 1 ]/H/I/C[0 1 1]}%
   user{/Subtype/Link/A<</Type/Action/S/URI/URI(#1)>>}%
  \relax
 }%
 \providecommand\@@endlink[0]{\pdfendlink}%
}%
\providecommand \url  [0]{\begingroup\@sanitize \@url }%
\providecommand \@url [1]{\endgroup\@href {#1}{\urlprefix}}%
\providecommand \urlprefix [0]{URL }%
\providecommand \Eprint[0]{\href }%
\@ifxundefined \urlstyle {%
  \providecommand \doi [1]{doi:\discretionary{}{}{}#1}%
}{%
  \providecommand \doi [0]{doi:\discretionary{}{}{}\begingroup
  \urlstyle{rm}\Url }%
}%
\providecommand \doibase [0]{http://dx.doi.org/}%
\providecommand \Doi[1]{\href{\doibase#1}}%
\providecommand \bibAnnote [3]{%
  \BibitemShut{#1}%
  \begin{quotation}\noindent
    \textsc{Key:}\ #2\\\textsc{Annotation:}\ #3%
  \end{quotation}%
}%
\providecommand \bibAnnoteFile [2]{%
  \IfFileExists{#2}{\bibAnnote {#1} {#2} {\input{#2}}}{}%
}%
\providecommand \typeout [0]{\immediate \write \m@ne }%
\providecommand \selectlanguage [0]{\@gobble}%
\providecommand \bibinfo [0]{\@secondoftwo}%
\providecommand \bibfield [0]{\@secondoftwo}%
\providecommand \translation [1]{[#1]}%
\providecommand \BibitemOpen[0]{}%
\providecommand \bibitemStop [0]{}%
\providecommand \bibitemNoStop [0]{.\EOS\space}%
\providecommand \EOS [0]{\spacefactor3000\relax}%
\providecommand \BibitemShut [1]{\csname bibitem#1\endcsname}%
\bibitem{Guth:1980}%
  \BibitemOpen
  \bibfield{author}{%
  \bibinfo {author} {\bibfnamefont{A.~H.}\ \bibnamefont{Guth}},\ }%
  \bibfield{journal}{%
  \Doi{10.1103/PhysRevD.23.347}{\bibinfo {journal} {Phys. Rev.}}\ }%
  \textbf{\bibinfo {volume} {D23}},\ \bibinfo {pages} {347} (\bibinfo {year}
  {1981})%
  \bibAnnoteFile{NoStop}{Guth:1980}%
\bibitem{La:1989za}%
  \BibitemOpen
  \bibfield{author}{%
  \bibinfo {author} {\bibfnamefont{D.}~\bibnamefont{La}}\ and\ \bibinfo
  {author} {\bibfnamefont{P.~J.}\ \bibnamefont{Steinhardt}},\ }%
  \bibfield{journal}{%
  \Doi{10.1103/PhysRevLett.62.376}{\bibinfo {journal} {Phys. Rev. Lett.}}\ }%
  \textbf{\bibinfo {volume} {62}},\ \bibinfo {pages} {376} (\bibinfo {year}
  {1989})%
  \bibAnnoteFile{NoStop}{La:1989za}%
\bibitem{Salopek:1988qh}%
  \BibitemOpen
  \bibfield{author}{%
  \bibinfo {author} {\bibfnamefont{D.~S.}\ \bibnamefont{Salopek}}, \bibinfo
  {author} {\bibfnamefont{J.~R.}\ \bibnamefont{Bond}},\ and\ \bibinfo {author}
  {\bibfnamefont{J.~M.}\ \bibnamefont{Bardeen}},\ }%
  \bibfield{journal}{%
  \Doi{10.1103/PhysRevD.40.1753}{\bibinfo {journal} {Phys. Rev.}}\ }%
  \textbf{\bibinfo {volume} {D40}},\ \bibinfo {pages} {1753} (\bibinfo {year}
  {1989})%
  \bibAnnoteFile{NoStop}{Salopek:1988qh}%
\bibitem{Futamase:1987ua}%
  \BibitemOpen
  \bibfield{author}{%
  \bibinfo {author} {\bibfnamefont{T.}~\bibnamefont{Futamase}}\ and\ \bibinfo
  {author} {\bibfnamefont{K.-i.}\ \bibnamefont{Maeda}},\ }%
  \bibfield{journal}{%
  \Doi{10.1103/PhysRevD.39.399}{\bibinfo {journal} {Phys. Rev.}}\ }%
  \textbf{\bibinfo {volume} {D39}},\ \bibinfo {pages} {399} (\bibinfo {year}
  {1989})%
  \bibAnnoteFile{NoStop}{Futamase:1987ua}%
\bibitem{Fakir:1990eg}%
  \BibitemOpen
  \bibfield{author}{%
  \bibinfo {author} {\bibfnamefont{R.}~\bibnamefont{Fakir}}\ and\ \bibinfo
  {author} {\bibfnamefont{W.~G.}\ \bibnamefont{Unruh}},\ }%
  \bibfield{journal}{%
  \Doi{10.1103/PhysRevD.41.1783}{\bibinfo {journal} {Phys. Rev.}}\ }%
  \textbf{\bibinfo {volume} {D41}},\ \bibinfo {pages} {1783} (\bibinfo {year}
  {1990})%
  \bibAnnoteFile{NoStop}{Fakir:1990eg}%
\bibitem{Bezrukov:2007ep}%
  \BibitemOpen
  \bibfield{author}{%
  \bibinfo {author} {\bibfnamefont{F.~L.}\ \bibnamefont{Bezrukov}}\ and\
  \bibinfo {author} {\bibfnamefont{M.}~\bibnamefont{Shaposhnikov}},\ }%
  \bibfield{journal}{%
  \Doi{10.1016/j.physletb.2007.11.072}{\bibinfo {journal} {Phys. Lett.}}\ }%
  \textbf{\bibinfo {volume} {B659}},\ \bibinfo {pages} {703} (\bibinfo {year}
  {2008}),\ \Eprint{http://arxiv.org/abs/0710.3755}{arXiv:0710.3755 [hep-th]}%
  \bibAnnoteFile{NoStop}{Bezrukov:2007ep}%
\bibitem{Komatsu:2010fb}%
  \BibitemOpen
  \bibfield{author}{%
  \bibinfo {author} {\bibfnamefont{E.}~\bibnamefont{Komatsu}} \emph{et~al.}}%
   (\bibinfo {year} {2010}),\
  \Eprint{http://arxiv.org/abs/1001.4538}{arXiv:1001.4538 [astro-ph.CO]}%
  \bibAnnoteFile{NoStop}{Komatsu:2010fb}%
\bibitem{Bardeen:1980kt}%
  \BibitemOpen
  \bibfield{author}{%
  \bibinfo {author} {\bibfnamefont{J.~M.}\ \bibnamefont{Bardeen}},\ }%
  \bibfield{journal}{%
  \Doi{10.1103/PhysRevD.22.1882}{\bibinfo {journal} {Phys. Rev.}}\ }%
  \textbf{\bibinfo {volume} {D22}},\ \bibinfo {pages} {1882} (\bibinfo {year}
  {1980})%
  \bibAnnoteFile{NoStop}{Bardeen:1980kt}%
\bibitem{Mukhanov:1981xt}%
  \BibitemOpen
  \bibfield{author}{%
  \bibinfo {author} {\bibfnamefont{V.~F.}\ \bibnamefont{Mukhanov}}\ and\
  \bibinfo {author} {\bibfnamefont{G.~V.}\ \bibnamefont{Chibisov}},\ }%
  \bibfield{journal}{%
  \bibinfo {journal} {JETP Lett.}\ }%
  \textbf{\bibinfo {volume} {33}},\ \bibinfo {pages} {532} (\bibinfo {year}
  {1981})%
  \bibAnnoteFile{NoStop}{Mukhanov:1981xt}%
\bibitem{Hawking:1982cz}%
  \BibitemOpen
  \bibfield{author}{%
  \bibinfo {author} {\bibfnamefont{S.~W.}\ \bibnamefont{Hawking}},\ }%
  \bibfield{journal}{%
  \Doi{10.1016/0370-2693(82)90373-2}{\bibinfo {journal} {Phys. Lett.}}\ }%
  \textbf{\bibinfo {volume} {B115}},\ \bibinfo {pages} {295} (\bibinfo {year}
  {1982})%
  \bibAnnoteFile{NoStop}{Hawking:1982cz}%
\bibitem{Starobinsky:1982ee}%
  \BibitemOpen
  \bibfield{author}{%
  \bibinfo {author} {\bibfnamefont{A.~A.}\ \bibnamefont{Starobinsky}},\ }%
  \bibfield{journal}{%
  \Doi{10.1016/0370-2693(82)90541-X}{\bibinfo {journal} {Phys. Lett.}}\ }%
  \textbf{\bibinfo {volume} {B117}},\ \bibinfo {pages} {175} (\bibinfo {year}
  {1982})%
  \bibAnnoteFile{NoStop}{Starobinsky:1982ee}%
\bibitem{Guth:1982ec}%
  \BibitemOpen
  \bibfield{author}{%
  \bibinfo {author} {\bibfnamefont{A.~H.}\ \bibnamefont{Guth}}\ and\ \bibinfo
  {author} {\bibfnamefont{S.~Y.}\ \bibnamefont{Pi}},\ }%
  \bibfield{journal}{%
  \Doi{10.1103/PhysRevLett.49.1110}{\bibinfo {journal} {Phys. Rev. Lett.}}\ }%
  \textbf{\bibinfo {volume} {49}},\ \bibinfo {pages} {1110} (\bibinfo {year}
  {1982})%
  \bibAnnoteFile{NoStop}{Guth:1982ec}%
\bibitem{Bardeen:1983qw}%
  \BibitemOpen
  \bibfield{author}{%
  \bibinfo {author} {\bibfnamefont{J.~M.}\ \bibnamefont{Bardeen}}, \bibinfo
  {author} {\bibfnamefont{P.~J.}\ \bibnamefont{Steinhardt}},\ and\ \bibinfo
  {author} {\bibfnamefont{M.~S.}\ \bibnamefont{Turner}},\ }%
  \bibfield{journal}{%
  \Doi{10.1103/PhysRevD.28.679}{\bibinfo {journal} {Phys. Rev.}}\ }%
  \textbf{\bibinfo {volume} {D28}},\ \bibinfo {pages} {679} (\bibinfo {year}
  {1983})%
  \bibAnnoteFile{NoStop}{Bardeen:1983qw}%
\bibitem{Mukhanov:1990me}%
  \BibitemOpen
  \bibfield{author}{%
  \bibinfo {author} {\bibfnamefont{V.~F.}\ \bibnamefont{Mukhanov}}, \bibinfo
  {author} {\bibfnamefont{H.~A.}\ \bibnamefont{Feldman}},\ and\ \bibinfo
  {author} {\bibfnamefont{R.~H.}\ \bibnamefont{Brandenberger}},\ }%
  \bibfield{journal}{%
  \Doi{10.1016/0370-1573(92)90044-Z}{\bibinfo {journal} {Phys. Rept.}}\ }%
  \textbf{\bibinfo {volume} {215}},\ \bibinfo {pages} {203} (\bibinfo {year}
  {1992})%
  \bibAnnoteFile{NoStop}{Mukhanov:1990me}%
\bibitem{Makino:1991sg}%
  \BibitemOpen
  \bibfield{author}{%
  \bibinfo {author} {\bibfnamefont{N.}~\bibnamefont{Makino}}\ and\ \bibinfo
  {author} {\bibfnamefont{M.}~\bibnamefont{Sasaki}},\ }%
  \bibfield{journal}{%
  \Doi{10.1143/PTP.86.103}{\bibinfo {journal} {Prog. Theor. Phys.}}\ }%
  \textbf{\bibinfo {volume} {86}},\ \bibinfo {pages} {103} (\bibinfo {year}
  {1991})%
  \bibAnnoteFile{NoStop}{Makino:1991sg}%
\bibitem{Fakir:1992cg}%
  \BibitemOpen
  \bibfield{author}{%
  \bibinfo {author} {\bibfnamefont{R.}~\bibnamefont{Fakir}}, \bibinfo {author}
  {\bibfnamefont{S.}~\bibnamefont{Habib}},\ and\ \bibinfo {author}
  {\bibfnamefont{W.}~\bibnamefont{Unruh}},\ }%
  \bibfield{journal}{%
  \Doi{10.1086/171591}{\bibinfo {journal} {Astrophys. J.}}\ }%
  \textbf{\bibinfo {volume} {394}},\ \bibinfo {pages} {396} (\bibinfo {year}
  {1992})%
  \bibAnnoteFile{NoStop}{Fakir:1992cg}%
\bibitem{Hwang:1996np}%
  \BibitemOpen
  \bibfield{author}{%
  \bibinfo {author} {\bibfnamefont{J.-c.}\ \bibnamefont{Hwang}},\ }%
  \bibfield{journal}{%
  \Doi{10.1088/0264-9381/14/7/029}{\bibinfo {journal} {Class. Quant. Grav.}}\
  }%
  \textbf{\bibinfo {volume} {14}},\ \bibinfo {pages} {1981} (\bibinfo {year}
  {1997}),\ \Eprint{http://arxiv.org/abs/gr-qc/9605024}{arXiv:gr-qc/9605024}%
  \bibAnnoteFile{NoStop}{Hwang:1996np}%
\bibitem{Hwang:1996xh}%
  \BibitemOpen
  \bibfield{author}{%
  \bibinfo {author} {\bibfnamefont{J.-c.}\ \bibnamefont{Hwang}}\ and\ \bibinfo
  {author} {\bibfnamefont{H.}~\bibnamefont{Noh}},\ }%
  \bibfield{journal}{%
  \Doi{10.1103/PhysRevD.54.1460}{\bibinfo {journal} {Phys. Rev.}}\ }%
  \textbf{\bibinfo {volume} {D54}},\ \bibinfo {pages} {1460} (\bibinfo {year}
  {1996})%
  \bibAnnoteFile{NoStop}{Hwang:1996xh}%
\bibitem{Prokopec:2010be}%
  \BibitemOpen
  \bibfield{author}{%
  \bibinfo {author} {\bibfnamefont{T.}~\bibnamefont{Prokopec}}\ and\ \bibinfo
  {author} {\bibfnamefont{G.}~\bibnamefont{Rigopoulos}},\ }%
  \bibfield{journal}{%
  \Doi{10.1103/PhysRevD.82.023529}{\bibinfo {journal} {Phys. Rev.}}\ }%
  \textbf{\bibinfo {volume} {D82}},\ \bibinfo {pages} {023529} (\bibinfo {year}
  {2010}),\ \Eprint{http://arxiv.org/abs/1004.0882}{arXiv:1004.0882 [gr-qc]}%
  \bibAnnoteFile{NoStop}{Prokopec:2010be}%
\bibitem{Chisholm1961469}%
  \BibitemOpen
  \bibfield{author}{%
  \bibinfo {author} {\bibfnamefont{J.~S.~R.}\ \bibnamefont{Chisholm}},\ }%
  \bibfield{journal}{%
  \Doi{10.1016/0029-5582(61)90106-7}{\bibinfo {journal} {Nuclear Physics}}\ }%
  \textbf{\bibinfo {volume} {26}},\ \bibinfo {pages} {469 } (\bibinfo {year}
  {1961})%
  \bibAnnoteFile{NoStop}{Chisholm1961469}%
\bibitem{Kamefuchi1961529}%
  \BibitemOpen
  \bibfield{author}{%
  \bibinfo {author} {\bibfnamefont{S.}~\bibnamefont{Kamefuchi}}, \bibinfo
  {author} {\bibfnamefont{L.}~\bibnamefont{O'Raifeartaigh}},\ and\ \bibinfo
  {author} {\bibfnamefont{A.}~\bibnamefont{Salam}},\ }%
  \bibfield{journal}{%
  \Doi{10.1016/0029-5582(61)91075-6}{\bibinfo {journal} {Nuclear Physics}}\ }%
  \textbf{\bibinfo {volume} {28}},\ \bibinfo {pages} {529 } (\bibinfo {year}
  {1961})%
  \bibAnnoteFile{NoStop}{Kamefuchi1961529}%
\bibitem{Bilandzic:2007nb}%
  \BibitemOpen
  \bibfield{author}{%
  \bibinfo {author} {\bibfnamefont{A.}~\bibnamefont{Bilandzic}}\ and\ \bibinfo
  {author} {\bibfnamefont{T.}~\bibnamefont{Prokopec}},\ }%
  \bibfield{journal}{%
  \Doi{10.1103/PhysRevD.76.103507}{\bibinfo {journal} {Phys. Rev.}}\ }%
  \textbf{\bibinfo {volume} {D76}},\ \bibinfo {pages} {103507} (\bibinfo {year}
  {2007}),\ \Eprint{http://arxiv.org/abs/0704.1905}{arXiv:0704.1905
  [astro-ph]}%
  \bibAnnoteFile{NoStop}{Bilandzic:2007nb}%
\bibitem{Arnowitt:1962hi}%
  \BibitemOpen
  \bibfield{author}{%
  \bibinfo {author} {\bibfnamefont{R.~L.}\ \bibnamefont{Arnowitt}}, \bibinfo
  {author} {\bibfnamefont{S.}~\bibnamefont{Deser}},\ and\ \bibinfo {author}
  {\bibfnamefont{C.~W.}\ \bibnamefont{Misner}}}%
   (\bibinfo {year} {1962}),\
  \Eprint{http://arxiv.org/abs/gr-qc/0405109}{arXiv:gr-qc/0405109}%
  \bibAnnoteFile{NoStop}{Arnowitt:1962hi}%
\bibitem{Maldacena:2002vr}%
  \BibitemOpen
  \bibfield{author}{%
  \bibinfo {author} {\bibfnamefont{J.~M.}\ \bibnamefont{Maldacena}},\ }%
  \bibfield{journal}{%
  \bibinfo {journal} {JHEP}\ }%
  \textbf{\bibinfo {volume} {05}},\ \bibinfo {pages} {013} (\bibinfo {year}
  {2003}),\
  \Eprint{http://arxiv.org/abs/astro-ph/0210603}{arXiv:astro-ph/0210603}%
  \bibAnnoteFile{NoStop}{Maldacena:2002vr}%
\bibitem{Feng:2010ya}%
  \BibitemOpen
  \bibfield{author}{%
  \bibinfo {author} {\bibfnamefont{C.-J.}\ \bibnamefont{Feng}}, \bibinfo
  {author} {\bibfnamefont{X.-Z.}\ \bibnamefont{Li}},\ and\ \bibinfo {author}
  {\bibfnamefont{E.~N.}\ \bibnamefont{Saridakis}},\ }%
  \bibfield{journal}{%
  \Doi{10.1103/PhysRevD.82.023526}{\bibinfo {journal} {Phys.Rev.}}\ }%
  \textbf{\bibinfo {volume} {D82}},\ \bibinfo {pages} {023526} (\bibinfo {year}
  {2010}),\ \Eprint{http://arxiv.org/abs/1004.1874}{arXiv:1004.1874
  [astro-ph.CO]}%
  \bibAnnoteFile{NoStop}{Feng:2010ya}%
\bibitem{Easson:2009wc}%
  \BibitemOpen
  \bibfield{author}{%
  \bibinfo {author} {\bibfnamefont{D.~A.}\ \bibnamefont{Easson}}, \bibinfo
  {author} {\bibfnamefont{S.}~\bibnamefont{Mukohyama}},\ and\ \bibinfo {author}
  {\bibfnamefont{B.~A.}\ \bibnamefont{Powell}},\ }%
  \bibfield{journal}{%
  \Doi{10.1103/PhysRevD.81.023512}{\bibinfo {journal} {Phys. Rev.}}\ }%
  \textbf{\bibinfo {volume} {D81}},\ \bibinfo {pages} {023512} (\bibinfo {year}
  {2010}),\ \Eprint{http://arxiv.org/abs/0910.1353}{arXiv:0910.1353
  [astro-ph.CO]}%
  \bibAnnoteFile{NoStop}{Easson:2009wc}%
\bibitem{Chiba:2008ia}%
  \BibitemOpen
  \bibfield{author}{%
  \bibinfo {author} {\bibfnamefont{T.}~\bibnamefont{Chiba}}\ and\ \bibinfo
  {author} {\bibfnamefont{M.}~\bibnamefont{Yamaguchi}},\ }%
  \bibfield{journal}{%
  \Doi{10.1088/1475-7516/2008/10/021}{\bibinfo {journal} {JCAP}}\ }%
  \textbf{\bibinfo {volume} {0810}},\ \bibinfo {pages} {021} (\bibinfo {year}
  {2008}),\ \Eprint{http://arxiv.org/abs/0807.4965}{arXiv:0807.4965
  [astro-ph]}%
  \bibAnnoteFile{NoStop}{Chiba:2008ia}%
\bibitem{Bezrukov:2008ej}%
  \BibitemOpen
  \bibfield{author}{%
  \bibinfo {author} {\bibfnamefont{F.~L.}\ \bibnamefont{Bezrukov}}, \bibinfo
  {author} {\bibfnamefont{A.}~\bibnamefont{Magnin}},\ and\ \bibinfo {author}
  {\bibfnamefont{M.}~\bibnamefont{Shaposhnikov}},\ }%
  \bibfield{journal}{%
  \Doi{10.1016/j.physletb.2009.03.035}{\bibinfo {journal} {Phys. Lett.}}\ }%
  \textbf{\bibinfo {volume} {B675}},\ \bibinfo {pages} {88} (\bibinfo {year}
  {2009}),\ \Eprint{http://arxiv.org/abs/0812.4950}{arXiv:0812.4950 [hep-ph]}%
  \bibAnnoteFile{NoStop}{Bezrukov:2008ej}%
\bibitem{DeSimone:2008ei}%
  \BibitemOpen
  \bibfield{author}{%
  \bibinfo {author} {\bibfnamefont{A.}~\bibnamefont{De~Simone}}, \bibinfo
  {author} {\bibfnamefont{M.~P.}\ \bibnamefont{Hertzberg}},\ and\ \bibinfo
  {author} {\bibfnamefont{F.}~\bibnamefont{Wilczek}},\ }%
  \bibfield{journal}{%
  \Doi{10.1016/j.physletb.2009.05.054}{\bibinfo {journal} {Phys. Lett.}}\ }%
  \textbf{\bibinfo {volume} {B678}},\ \bibinfo {pages} {1} (\bibinfo {year}
  {2009}),\ \Eprint{http://arxiv.org/abs/0812.4946}{arXiv:0812.4946 [hep-ph]}%
  \bibAnnoteFile{NoStop}{DeSimone:2008ei}%
\bibitem{Bezrukov:2009db}%
  \BibitemOpen
  \bibfield{author}{%
  \bibinfo {author} {\bibfnamefont{F.}~\bibnamefont{Bezrukov}}\ and\ \bibinfo
  {author} {\bibfnamefont{M.}~\bibnamefont{Shaposhnikov}},\ }%
  \bibfield{journal}{%
  \Doi{10.1088/1126-6708/2009/07/089}{\bibinfo {journal} {JHEP}}\ }%
  \textbf{\bibinfo {volume} {07}},\ \bibinfo {pages} {089} (\bibinfo {year}
  {2009}),\ \Eprint{http://arxiv.org/abs/0904.1537}{arXiv:0904.1537 [hep-ph]}%
  \bibAnnoteFile{NoStop}{Bezrukov:2009db}%
\bibitem{Barvinsky:2008ia}%
  \BibitemOpen
  \bibfield{author}{%
  \bibinfo {author} {\bibfnamefont{A.~O.}\ \bibnamefont{Barvinsky}}, \bibinfo
  {author} {\bibfnamefont{A.~Y.}\ \bibnamefont{Kamenshchik}},\ and\ \bibinfo
  {author} {\bibfnamefont{A.~A.}\ \bibnamefont{Starobinsky}},\ }%
  \bibfield{journal}{%
  \Doi{10.1088/1475-7516/2008/11/021}{\bibinfo {journal} {JCAP}}\ }%
  \textbf{\bibinfo {volume} {0811}},\ \bibinfo {pages} {021} (\bibinfo {year}
  {2008}),\ \Eprint{http://arxiv.org/abs/0809.2104}{arXiv:0809.2104 [hep-ph]}%
  \bibAnnoteFile{NoStop}{Barvinsky:2008ia}%
\bibitem{Barvinsky:2009fy}%
  \BibitemOpen
  \bibfield{author}{%
  \bibinfo {author} {\bibfnamefont{A.~O.}\ \bibnamefont{Barvinsky}}, \bibinfo
  {author} {\bibfnamefont{A.~Y.}\ \bibnamefont{Kamenshchik}}, \bibinfo {author}
  {\bibfnamefont{C.}~\bibnamefont{Kiefer}}, \bibinfo {author}
  {\bibfnamefont{A.~A.}\ \bibnamefont{Starobinsky}},\ and\ \bibinfo {author}
  {\bibfnamefont{C.}~\bibnamefont{Steinwachs}},\ }%
  \bibfield{journal}{%
  \Doi{10.1088/1475-7516/2009/12/003}{\bibinfo {journal} {JCAP}}\ }%
  \textbf{\bibinfo {volume} {0912}},\ \bibinfo {pages} {003} (\bibinfo {year}
  {2009}),\ \Eprint{http://arxiv.org/abs/0904.1698}{arXiv:0904.1698 [hep-ph]}%
  \bibAnnoteFile{NoStop}{Barvinsky:2009fy}%
\bibitem{Barvinsky:2009ii}%
  \BibitemOpen
  \bibfield{author}{%
  \bibinfo {author} {\bibfnamefont{A.~O.}\ \bibnamefont{Barvinsky}}, \bibinfo
  {author} {\bibfnamefont{A.~Y.}\ \bibnamefont{Kamenshchik}}, \bibinfo {author}
  {\bibfnamefont{C.}~\bibnamefont{Kiefer}}, \bibinfo {author}
  {\bibfnamefont{A.~A.}\ \bibnamefont{Starobinsky}},\ and\ \bibinfo {author}
  {\bibfnamefont{C.~F.}\ \bibnamefont{Steinwachs}}}%
   (\bibinfo {year} {2009}),\
  \Eprint{http://arxiv.org/abs/0910.1041}{arXiv:0910.1041 [hep-ph]}%
  \bibAnnoteFile{NoStop}{Barvinsky:2009ii}%
\bibitem{Burgess:2009ea}%
  \BibitemOpen
  \bibfield{author}{%
  \bibinfo {author} {\bibfnamefont{C.~P.}\ \bibnamefont{Burgess}}, \bibinfo
  {author} {\bibfnamefont{H.~M.}\ \bibnamefont{Lee}},\ and\ \bibinfo {author}
  {\bibfnamefont{M.}~\bibnamefont{Trott}},\ }%
  \bibfield{journal}{%
  \Doi{10.1088/1126-6708/2009/09/103}{\bibinfo {journal} {JHEP}}\ }%
  \textbf{\bibinfo {volume} {09}},\ \bibinfo {pages} {103} (\bibinfo {year}
  {2009}),\ \Eprint{http://arxiv.org/abs/0902.4465}{arXiv:0902.4465 [hep-ph]}%
  \bibAnnoteFile{NoStop}{Burgess:2009ea}%
\bibitem{Lerner:2009na}%
  \BibitemOpen
  \bibfield{author}{%
  \bibinfo {author} {\bibfnamefont{R.~N.}\ \bibnamefont{Lerner}}\ and\ \bibinfo
  {author} {\bibfnamefont{J.}~\bibnamefont{McDonald}},\ }%
  \bibfield{journal}{%
  \Doi{10.1088/1475-7516/2010/04/015}{\bibinfo {journal} {JCAP}}\ }%
  \textbf{\bibinfo {volume} {1004}},\ \bibinfo {pages} {015} (\bibinfo {year}
  {2010}),\ \Eprint{http://arxiv.org/abs/0912.5463}{arXiv:0912.5463 [hep-ph]}%
  \bibAnnoteFile{NoStop}{Lerner:2009na}%
\bibitem{Hertzberg:2010dc}%
  \BibitemOpen
  \bibfield{author}{%
  \bibinfo {author} {\bibfnamefont{M.~P.}\ \bibnamefont{Hertzberg}}}%
   (\bibinfo {year} {2010}),\
  \Eprint{http://arxiv.org/abs/1002.2995}{arXiv:1002.2995 [hep-ph]}%
  \bibAnnoteFile{NoStop}{Hertzberg:2010dc}%
\bibitem{Burgess:2010zq}%
  \BibitemOpen
  \bibfield{author}{%
  \bibinfo {author} {\bibfnamefont{C.~P.}\ \bibnamefont{Burgess}}, \bibinfo
  {author} {\bibfnamefont{H.~M.}\ \bibnamefont{Lee}},\ and\ \bibinfo {author}
  {\bibfnamefont{M.}~\bibnamefont{Trott}},\ }%
  \bibfield{journal}{%
  \Doi{10.1007/JHEP07(2010)007}{\bibinfo {journal} {JHEP}}\ }%
  \textbf{\bibinfo {volume} {07}},\ \bibinfo {pages} {007} (\bibinfo {year}
  {2010}),\ \Eprint{http://arxiv.org/abs/1002.2730}{arXiv:1002.2730 [hep-ph]}%
  \bibAnnoteFile{NoStop}{Burgess:2010zq}%
\bibitem{Kaiser:2010ps}%
  \BibitemOpen
  \bibfield{author}{%
  \bibinfo {author} {\bibfnamefont{D.~I.}\ \bibnamefont{Kaiser}},\ }%
  \bibfield{journal}{%
  \Doi{10.1103/PhysRevD.81.084044}{\bibinfo {journal} {Phys.Rev.}}\ }%
  \textbf{\bibinfo {volume} {D81}},\ \bibinfo {pages} {084044} (\bibinfo {year}
  {2010}),\ \Eprint{http://arxiv.org/abs/1003.1159}{arXiv:1003.1159 [gr-qc]}%
  \bibAnnoteFile{NoStop}{Kaiser:2010ps}%
\bibitem{Bezrukov:2010jz}%
  \BibitemOpen
  \bibfield{author}{%
  \bibinfo {author} {\bibfnamefont{F.}~\bibnamefont{Bezrukov}}, \bibinfo
  {author} {\bibfnamefont{A.}~\bibnamefont{Magnin}}, \bibinfo {author}
  {\bibfnamefont{M.}~\bibnamefont{Shaposhnikov}},\ and\ \bibinfo {author}
  {\bibfnamefont{S.}~\bibnamefont{Sibiryakov}}}%
   (\bibinfo {year} {2010}),\
  \Eprint{http://arxiv.org/abs/1008.5157}{arXiv:1008.5157 [hep-ph]}%
  \bibAnnoteFile{NoStop}{Bezrukov:2010jz}%
\bibitem{Janssen:2009pb}%
  \BibitemOpen
  \bibfield{author}{%
  \bibinfo {author} {\bibfnamefont{T.~M.}\ \bibnamefont{Janssen}}, \bibinfo
  {author} {\bibfnamefont{S.~P.}\ \bibnamefont{Miao}}, \bibinfo {author}
  {\bibfnamefont{T.}~\bibnamefont{Prokopec}},\ and\ \bibinfo {author}
  {\bibfnamefont{R.~P.}\ \bibnamefont{Woodard}},\ }%
  \bibfield{journal}{%
  \Doi{10.1088/1475-7516/2009/05/003}{\bibinfo {journal} {JCAP}}\ }%
  \textbf{\bibinfo {volume} {0905}},\ \bibinfo {pages} {003} (\bibinfo {year}
  {2009}),\ \Eprint{http://arxiv.org/abs/0904.1151}{arXiv:0904.1151 [gr-qc]}%
  \bibAnnoteFile{NoStop}{Janssen:2009pb}%
\bibitem{Sloth:2006az}%
  \BibitemOpen
  \bibfield{author}{%
  \bibinfo {author} {\bibfnamefont{M.~S.}\ \bibnamefont{Sloth}},\ }%
  \bibfield{journal}{%
  \Doi{10.1016/j.nuclphysb.2006.04.029}{\bibinfo {journal} {Nucl. Phys.}}\ }%
  \textbf{\bibinfo {volume} {B748}},\ \bibinfo {pages} {149} (\bibinfo {year}
  {2006}),\
  \Eprint{http://arxiv.org/abs/astro-ph/0604488}{arXiv:astro-ph/0604488}%
  \bibAnnoteFile{NoStop}{Sloth:2006az}%
\bibitem{Seery:2007wf}%
  \BibitemOpen
  \bibfield{author}{%
  \bibinfo {author} {\bibfnamefont{D.}~\bibnamefont{Seery}},\ }%
  \bibfield{journal}{%
  \Doi{10.1088/1475-7516/2008/02/006}{\bibinfo {journal} {JCAP}}\ }%
  \textbf{\bibinfo {volume} {0802}},\ \bibinfo {pages} {006} (\bibinfo {year}
  {2008}),\ \Eprint{http://arxiv.org/abs/0707.3378}{arXiv:0707.3378
  [astro-ph]}%
  \bibAnnoteFile{NoStop}{Seery:2007wf}%
\bibitem{Burgess:2009bs}%
  \BibitemOpen
  \bibfield{author}{%
  \bibinfo {author} {\bibfnamefont{C.~P.}\ \bibnamefont{Burgess}}, \bibinfo
  {author} {\bibfnamefont{L.}~\bibnamefont{Leblond}}, \bibinfo {author}
  {\bibfnamefont{R.}~\bibnamefont{Holman}},\ and\ \bibinfo {author}
  {\bibfnamefont{S.}~\bibnamefont{Shandera}},\ }%
  \bibfield{journal}{%
  \Doi{10.1088/1475-7516/2010/03/033}{\bibinfo {journal} {JCAP}}\ }%
  \textbf{\bibinfo {volume} {1003}},\ \bibinfo {pages} {033} (\bibinfo {year}
  {2010}),\ \Eprint{http://arxiv.org/abs/0912.1608}{arXiv:0912.1608 [hep-th]}%
  \bibAnnoteFile{NoStop}{Burgess:2009bs}%
\end{thebibliography}%

\end{document}